\documentclass[epj]{svjour}
\usepackage{graphicx}
\usepackage{amssymb}
\usepackage{bm}
\usepackage{amsmath} 
\usepackage{nicefrac}
\newcommand{\SO}{${}^1\textrm{S}_0\;$}
\newcommand{\SD}{${}^3\textrm{SD}_1\;$}

\begin{document}
\title{ Low-density homogeneous symmetric nuclear matter:
               disclosing dinucleons in coexisting phases }
\author{Hugo F. Arellano\inst{1,2} \and Jean-Paul Delaroche\inst{1}
}                     
%
%
\institute{
CEA, DAM, DIF, F-91297 Arpajon, France \and
Department of Physics - FCFM, University of Chile,
          Av. Blanco Encalada 2008, Santiago, Chile
}
\date{Received: date / Revised version: date}
%
\abstract{
The effect of \emph{in-medium} dinucleon bound states on
self-consistent single-particle fields in Brueckner, Bethe and Goldstone
theory is investigated in symmetric nuclear matter at zero temperature.
To this end, dinucleon bound state occurences in the \SO and
\SD channels are explicitly accounted for
--within the continuous choice for the auxiliary fields--
while imposing self-consistency in Brueckner-Hartree-Fock
approximation calculations.
Searches are carried out at Fermi momenta in the
range $0<k_F\leq1.75$~fm$^{-1}$, using the Argonne $v_{18}$ bare
nucleon-nucleon potential without resorting to the effective
mass approximation.
As a result, two distinct solutions meeting the self-consistency 
requirement are found with overlapping domains in the interval
$0.130\;\textrm{fm}^{-1} \leq k_F \leq 0.285\;\textrm{fm}^{-1}$,
corresponding to mass densities 
between $10^{11.4}$ and $10^{12.4}$~g~cm$^{-3}$.
Effective masses as high as three times the nucleon mass are found in
the coexistence domain.
The emergence of superfluidity in relationship with BCS pairing gap 
solutions is discussed.
\PACS{
      {21.65.-f}{Nuclear matter}\and
      {21.45.Bc}{Two-nucleon system}\and
      {21.65.Mn}{Equations of state of nuclear matter}
}
} 
\titlerunning{Low-density homogeneous symmetric nuclear matter...}
\authorrunning{Arellano \& Delaroche}

\maketitle
\section{Introduction}
\label{intro}
Dinucleon formation, pairing, clustering and condensation in 
nuclear media have been subjects of considerable research 
during the past few decades motivated by their role 
in heavy-ion collisions, collapsing stars, weakly bound nuclear systems 
and critical phenomena \cite{Broglia13,Sedrakian06}.
Even in their simplest conceptions, nuclear matter models may become 
exceedingly complex systems featuring strongly-interacting Fermions
over a broad range of conditions.
Densities may vary from highly diluted systems, 
as in nuclear halos or envelopes of core-collapse supernovae, 
up to a few times the saturation density, as in the core of neutron stars. 
The isospin asymmetry can also vary from zero, 
as in symmetric nuclear matter, 
up to unity, as in pure neutron matter thought to prevail
in the interior of neutron stars.
Temperature can also be subject to wide variations
from zero up to a few hundreds of MeV.

Among the various approaches to study interacting nucleons systems,
the Brueckner-Hartree-Fock (BHF)
approximation can be regarded as
a reasonable starting point to investigate homogeneous
infinite nuclear matter \cite{Li06,Gogelin08}.
This approximation embodies important information 
upon which higher order corrections can be incorporated \cite{Dickh08}.
However, it has long been recognized that bound states in BHF approximation 
signal the onset of superfluidity,
phenomenon which needs to be investigated within a much 
broader framework \cite{Muther2005}.
Beyond this limitation, the question remains as to whether it is feasible to
find self-consistent single-particle (sp) fields in infinite nuclear matter
within BHF, accounting \emph{simultaneously} for dinucleon bound states.
In this work we address this issue by providing an explicit 
account for dinucleon bound states in the ${}^1\textrm{S}_0$ and \SD
channels, expressed as isolated poles of the $g$ matrix 
on the real axis below the continuum threshold,
while seeking self-consistency in the sp potential.
We have avoided the use of the effective mass approximation to 
represent the sp fields, aiming to rule out potential sources of 
ambiguities in our findings.
Additionally, the sp momentum distribution follows a normal Fermi 
distribution, namely 
$n(k)=\Theta(k_F-k)$, with $k_F$ the Fermi momentum.
As a consequence we obtain two distinct solutions 
satisfying the self-consistency requirement in the range 
$0.13\,\textrm{fm}^{-1}\lesssim k_F\lesssim \,0.29\,\textrm{fm}^{-1}$,
\emph{i.e.} matter densities between 
$1.5\times 10^{-4}$ and $2\times 10^{-3}$~fm$^{-3}$, corresponding to
mass densities of
$10^{11.4}$ and $10^{12.4}$~g~cm$^{-3}$, respectively.
To investigate and resolve these features, special measures were
needed to control singularities implied by bound states.
Considering that at very low densities hole-hole ladder contributions become
weak \cite{Ramos89} due to the limited phase space they act upon,
the results we obtain for nuclear densities below 
$10^{-3}\,\textrm{fm}^{-3}$ based only on particle-particle ({pp})
contributions become reasonably justified within BHF alone,
namely as long as effects associated with superfluidity are neglected.
In any case, taking the present results as a baseline,
the inclusion of hole-hole propagation beyond Brueckner
ladder diagrams would need to be reassessed in another work.

Nuclear matter properties have been extensively investigated in the 
framework of the Brueckner-Bethe-Goldstone (BBG) theory based on the 
hole-line expansion for the ground state energy \cite{Baldo99}.
Here the Goldstone diagrams are grouped according to their number 
of hole lines, with each group summed up separately.
The summation of the two-hole-line diagrams yields the usual 
BHF approximation, where the two-body 
scattering matrix in nuclear matter is calculated self-consistently 
with the sp energy spectrum.
Although the sp potential is introduced only as an auxiliary quantity,
its choice conditions the rate of convergence of the expansion for 
the binding energy.
In ref. \cite{Song98} it has been reported that the continuous
choice \cite{Lejeune78,Hufner72,Sartor83}
for the auxiliary potential provides better convergence
over the so called \emph{standard choice}, where the sp potential 
is set to zero above the Fermi energy. 
Hence, we shall use the continuous choice to investigate the role 
of dinucleon bound states during the search for
self-consistent sp potentials.

This paper is organized as follows.
In sec. \ref{theory} we outline the theoretical background upon which 
we address self-consistency in nuclear matter within the BHF approximation.
In sec. \ref{calcul} we succinctly describe special measures 
aimed to control numerical instabilities, particularly the treatment 
of dinucleon bound-state poles of the $g$ matrix on the real axis.
We also address the existence of multiple roots in the energy 
denominator of the propagator and describe procedures to 
handle sharp integrands.
In sec. \ref{results} we present and discuss solutions for the 
self-consistent sp fields, associated nuclear matter binding and 
effective masses.
We also discuss spatial properties of dinucleons
as well as some immediate implications of the sp fields 
when applied in the context of pairing 
gap equations for superfluid states. 
In sec. \ref{fin} we present a summary of the work, 
draw its main conclusions and outlook.

\section{Framework}
\label{theory}
In BBG theory for symmetric nuclear matter the $g$ matrix depends
on the density of the medium, characterized by the Fermi momentum $k_F$,
and a starting energy $\omega$.
To lowest order in the BHF approximation for nuclear matter 
in normal state, 
when only two-body correlations are taken into account,
the $g$ matrix satisfies
\begin{equation}
\label{bbg0}
g(\omega)=v+v\,\frac{Q}{\omega+i\eta-\hat h_1-\hat h_2}\,g(\omega)\,,
\end{equation}
with $v$ the bare interaction between nucleons,
$\hat h_{i}$ the sp energy of nucleon $i$ ($i=1,2$), 
and $Q$ the Pauli blocking operator which 
in normal BHF takes the form
\[
Q|\boldsymbol{ p\, k}\rangle =
\Theta(p-k_F)\Theta(k-k_F) |\boldsymbol{ p\, k}\rangle \;.
\]
The solution to eq.~(\ref{bbg0}) enables the evaluation of 
the mass operator
\begin{equation}
\label{mass}
M(k;E)=\sum_{\mid\boldsymbol p\mid\leq k_F}
\langle \textstyle{\frac{1}{2}}(\boldsymbol k-\boldsymbol p)
| g_{\boldsymbol K}(E+e_p) |
\textstyle{\frac{1}{2}}(\boldsymbol k-\boldsymbol p)\rangle\;,
\end{equation}
where  $\boldsymbol K$ is the total momentum of the interacting pair,
$\boldsymbol K = \boldsymbol k+\boldsymbol p$, and
\begin{equation}
\label{esp}
e_p=\frac{p^2}{2m} + U(p)\,,
\end{equation}
the sp energy defined in terms of an auxiliary field $U$.
The nucleon mass $m$ is taken as the average of proton and neutron masses. 
In the BHF approximation the sp potential is given by the
on-shell mass operator,
\begin{equation}
\label{usp}
U(k)=\textsf{Re}\, M(k;e_k)\;,
\end{equation}
self-consistency requirement which can be achieved iteratively.
In the continuous choice this condition is imposed at all 
momenta $k$ \cite{Baldo00}.

In momentum representation the BHF equation takes the explicit form
\begin{align}
\label{bbgq}
\langle\boldsymbol\kappa'|g_{\boldsymbol K}(\omega)|
\boldsymbol\kappa\rangle
 =&
\langle\boldsymbol\kappa'|v\,|\boldsymbol\kappa\rangle
+
\int d{\boldsymbol q}\,
\langle{\boldsymbol\kappa}'|v\,|{\boldsymbol q}\rangle\times \nonumber\\
& \frac{\Theta(k_{+}-k_F)\Theta(k_{-}-k_F)}
{\omega+i\,\eta-\frac{k_{+}^2}{2m}-\frac{k_{-}^2}{2m}-\Sigma}\,
\langle{\boldsymbol q}|g_{\boldsymbol K}(\omega)|
\boldsymbol\kappa\rangle\,,
\end{align}
where $\Sigma$ accounts for the particle-particle  ({pp}) potential 
\begin{equation}
\label{sigmau}
\Sigma(K,q;x)\equiv U(k_{+})+U(k_{-})\;,
\end{equation}
with 
\begin{equation}
\label{kpm}
k_{\pm}^2 = \frac{K^2}{4}+q^2\pm qKx\,.
\end{equation}
Here $x=\hat{\boldsymbol K}\cdot\hat{\boldsymbol q}$, 
corresponding to the cosine of
the angle between ${\boldsymbol K}$ and ${\boldsymbol q}$,
with ${\boldsymbol q}$ the relative momentum of intermediate states.

\section{Methods}
\label{calcul}
A standard procedure to solve the non-linear system of 
eqs. (\ref{bbg0}-\ref{usp}) is by iterative feed-backing,
where the auxiliary field $U$ is initially unknown.
Each cycle begins with the definition of an initial guess 
for $U(k)$, which allows solving eq.~(\ref{bbg0}) 
for any $\omega$ and $K$.
The series of cycles may be started with $U_0(k)=0$, 
to obtain all $g$-matrix elements needed for the evaluation 
of $M$ in eq.~(\ref{mass}). 
The real part of $M$ defines a new sp field, $U_1$, 
which becomes the guess for a new cycle. 
The procedure is repeated until differences between consecutive solutions
for $U$ meet some convergence criterion. 
Although this self-consistent scheme works well at normal densities,
convergence becomes more difficult as $k_F$ diminishes 
below $\sim$0.8~fm${}^{-1}$, feature manifested by instabilities in the
evaluation of the on-shell mass operator.
To attenuate these instabilities we cite 
refs. \cite{Aguayo08}, for example, 
where self-consistency at low densities was reached by reducing 
the number of mesh points for the Fermi-motion 
integral in eq.~(\ref{mass}).
With such numerical compromise it was possible to attenuate 
sudden fluctuations of the integrand, 
which at the time of the work had no identifiable cause.
These solutions are used to calculate
nucleon-nucleon (\emph{NN}) effective interaction,  
\emph{i.e.} off-shell $g$ matrix, 
to evaluate momentum-space optical model 
potentials for nucleon scattering off nuclei.

Aiming to obtain genuine self-consistent solutions for the sp mean 
fields over a wide density regime,
particularly at low densities,
a refinement of numerical techniques becomes crucial
to control instabilities during the iterative process.
In the following we describe the most relevant constructions 
toward this end.

\subsection{Multiple roots}
To solve eq.~(\ref{bbgq}) we take the angular average
of the energy denominator,
\begin{equation}
\label{uaveraged}
\bar\Sigma(K,q)=\frac{1}{\Delta}\int_0^\Delta [ U(k_{+})+U(k_{-}) ]\,dx\,,
\end{equation}
where
$\Delta=\Delta(K,q)=\min\{1,\max[0,(K^2/4+q^2-k_F^2)/qK]\}$.
Additionally, if the Pauli blocking operator is angle-averaged the
multi-channel coupling among different total angular momentum states 
become disentangled. 
In ref. \cite{Arellano11a} this is referred to as ratio-of-average 
approximation,
which in this study constitutes a starting point.
After partial wave expansion, eq.~(\ref{bbg0}) for uncoupled 
channels reads
\begin{align}
\label{ls_uncoupled}
g(k',k;\omega) =&
v(k',k) +
\frac{2}{\pi}\int_0^\infty q^2\,dq\,v(k',q) \times\nonumber\\
& \frac{\Delta(K,q)}{\omega+i\eta-E(K,q)}\,
g(q,k;\omega)\;,
\end{align}
where
\begin{equation}
\label{epair}
E(K,q)=\frac{K^2}{4m}+\frac{q^2}{m}+\bar\Sigma(K,q)\;.
\end{equation}
A similar equation is obtained in the case of coupled states.
The above integral equation can be solved numerically with the use of 
numerical quadrature and matrix inversion. 
However, caution is called for in the treatment of the zeros 
of the energy denominator.
Indeed, the fact that $e_k=k^2/2m+U(k)$ is a growing function in $k$ 
does not necessarily imply the same trend for the 
two-particle energy $E(K,q)$.
Hence, more than one root in the energy denominator of
eq.~(\ref{ls_uncoupled}) cannot be ruled out.

To clarify this point consider the angle-averaged 
potential $\bar\Sigma(K,q)$ in eq.~(\ref{uaveraged}) evaluated  
using a two-point Gaussian quadrature. Thus
\begin{equation}
\label{egauss}
E(K,q)\approx \frac{K^2}{4m}+\frac{q^2}{m}+U(p_{+}) + U(p_{-})\;,
\end{equation}
where $p_{\pm}=(K^2/4+q^2\pm Kq/\sqrt{3})^{1/2}$.
Single zeros (or none) in $E(K,q)-\omega$ are guaranteed if
$E(K,q)$ is an ever growing function in $q$, namely
$\partial E(K,q)/\partial q>0$. Hence,
\begin{equation}
\label{ineq}
\frac{2q}{m} + 
q \left [ 
\frac{U'(p_{+})}{p_{+}} +
\frac{U'(p_{-})}{p_{-}}
\right ] +
\frac{K}{2\sqrt{3}} \left [ 
\frac{U'(p_{+})}{p_{+}} -
\frac{U'(p_{-})}{p_{-}}
\right ]
>0\;,
\end{equation} 
where $U'$ stands for derivative of $U$ with respect to its argument.
Assuming $U'(k)$ positive for all $k$, 
the validity of the above inequality depends on the sign of the third 
term. Only if $U(k)$ is quadratic in $k$ the above inequality is met,
which is not the case in actual solutions.

Therefore, 
even if the genuine sp self-consistent fields result as monotonic 
functions in the momentum space, 
multiple roots cannot be ruled out in the energy 
denominator of the propagator.
This aspect, in the context of the integral equation for the $g$ matrix,
can be handled as outlined in Appendix \ref{multiple}.

\subsection{Fermi-motion integrals}
The evaluation of the sp fields, $U(k)=\textsf{Re}\,M(k,e_k)$, 
involves the summation of on-shell $g$-matrix elements while keeping
the momentum of one of the particles below the Fermi surface.
Explicitly, 
\begin{equation}
\label{FermiMotion}
M(k,e_k)=
\sum_{\alpha} n_\alpha \int_0^{k_F}p^2 dp\,\int_{-1}^{1}du\;
g^{\alpha}_{{\boldsymbol k}+{\boldsymbol p}}
(
\textstyle{\frac{|{\boldsymbol k}-{\boldsymbol p}|}{2}},
\textstyle{\frac{|{\boldsymbol k}-{\boldsymbol p}|}{2}};\omega
)\;.
\end{equation}
Here $u=\hat{\boldsymbol k}\cdot\hat{\boldsymbol p}$, 
the energy $\omega$ is evaluated on-shell ($\omega= e_k+e_p$),
$\alpha$ denotes spin, isospin and angular momentum states,
and $n_\alpha$ accounts for their degeneracy and for geometric factors.

\begin{figure}
\begin{center}
\resizebox{0.3\textwidth}{!}{%
\includegraphics[clip=true] {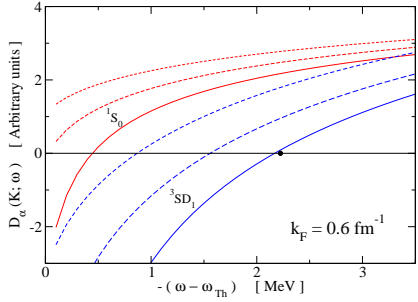}
}
\end{center}
\caption{{\protect
\label{determinant}
Determinant $D_\alpha(K,\omega)$
as a function of $\omega$ 
for ${}^1\textrm{S}_0$ (thin curves) and \SD (thick curves) channels,
at $k_F=0.6$~fm${}^{-1}$.
Solid, long- and short-dashed curves indicate
$K=0$, 0.05 and 0.1~fm${}^{-1}$, respectively.
The circle at 2.224~MeV denotes 
deuteron binding in free space.
}}
\end{figure}

For most \emph{NN} states the integration within the 
Fermi sphere involves well behaved integrands, 
as given by the on-shell $g$.
However, extra measures becomes unavoidable when considering 
the ${}^1\textrm{S}_0$ and \SD channels.
This is because during the evaluation of the momentum integral 
in eq.~(\ref{FermiMotion}), exceedingly large $g$-matrix elements
may occur,
stemming from \emph{NN} bound states in these S and D channels.
Indeed, any matrix element with starting energy near these bound states
will lead to large contributions.
To illustrate the occurrence of these states, let us define
\begin{equation}
\label{det}
D_\alpha(K;\omega)\equiv\det [1-v_\alpha \Lambda_K(\omega)]\;,
\end{equation}
with $\Lambda_K(\omega)$ the {pp} propagator in eq.~(\ref{ls_uncoupled})
and $\alpha$ denoting a particular \emph{NN} channel.
In fig.~\ref{determinant} we plot $D_\alpha$ 
for ${}^1\textrm{S}_0$ (thin curves) and \SD (thick curves) channels
as a function of $(\omega-\omega_{\textrm{th}})$, 
with $\omega_{\textrm{th}}$ the
threshold energy allowed by Pauli blocking. 
For this example we have chosen $k_F=0.6$~fm${}^{-1}$.
Solid, long- and short-dashed curves correspond to total pair momenta
$K=0$, 0.05 and 0.1~fm${}^{-1}$, respectively.
A singularity of $g_K(\omega)$ in channel $\alpha$ is identified by 
the zero of $D_\alpha(K;\omega)$.
As observed, 
the deuteron channel exhibits bound states in all three cases, 
while for the ${}^1\textrm{S}_0$ channel they occur over a 
narrower interval in $\omega$.
Hence, if during the evaluation of the $dp\,du$ integrals in 
eq.~(\ref{FermiMotion}) a configuration of total momentum $K$
and starting energy $\omega$ happens to be near (or at) one of 
these singular points,
large (or undefined) contributions become unavoidable.
This feature becomes a major source of numerical instabilities 
if not handled properly.

\begin{figure}
\begin{center}
\includegraphics[width=0.6\linewidth,clip=true] {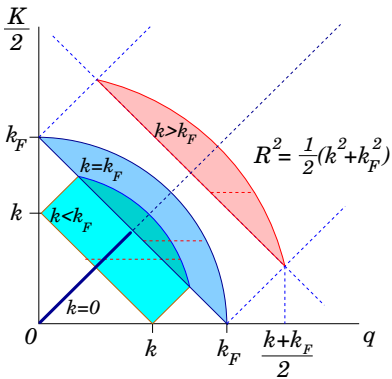}
\end{center}
\caption{{\protect
\label{FermiRegion}
 Domains of integration in the $(\textstyle{\frac12}K,q)$ plane
for different values of $k$ relative to $k_F$.
The radii $R$ of the arcs satisfy $R^2=(k^2+k_F^2)/2$. 
The slopes of the straight lines are $\pm 1$.
}}
\end{figure}

To evaluate the mass operator we use the
alternative set of variables
$q$ and $K$ defined by \cite{Haftel70}
\begin{subequations}
\label{Kq}
\begin{align}
q&=\textstyle{\frac12}(k^2+p^2-2kpu)^{1/2}\;,\label{Kq1}\\
K&=(k^2+p^2+2kpu)^{1/2}\;.\label{Kq2}
\end{align}
\end{subequations}
Thus, $p^2dp\,du = (2Kq/k) dK\,dq$.
Therefore the on-shell mass operator, eq.~(\ref{FermiMotion}),
is now expressed as
\begin{equation}
\label{MassOperator}
M(k,e_k)=\sum_{\alpha} n_\alpha 
\int_{K_{\textrm{min}}}^{K_{\textrm{max}}} K\,dK
\int_{q_{\textrm{min}}}^{q_{\textrm{max}}} \frac{2q\,dq}{k}\,
g^{\alpha}_{K} (q,q;\omega).
\end{equation}
To be consistent with the angle-averaged {pp} energy,
the starting energy $\omega=e_p+e_k$ is evaluated according to 
eq.~(\ref{epair}).  With this set of integration variables, 
for given $k_F$, $k$ and $K$, 
the limits of integration on $q$ become
\begin{align*}
\label{qLimits}
q_{\textrm{min}}&=|k-\textstyle{\frac12} K|\;, \\
q_{\textrm{max}}&=\left\{
\begin{array}{lc}
\min\{k+\frac12,({R^2-\frac14 K^2})^{1/2}\} 
\qquad  & \textrm{for $k\leq k_F$}\,, \\
({R^2-\frac14 K^2})^{1/2} & \textrm{for $k> k_F$}\;,
\end{array}
\right. 
\end{align*}
where $R^2=\textstyle{\frac12}\left ( k^2+k_F^2\right )$.
In the case of $K$, the integration ranges from
$K_{\textrm{min}}=\max\{ 0,k-k_F\}$,
up to $K_{\textrm{max}}=k+k_F$.
In fig.~\ref{FermiRegion} we illustrate the four possible shapes 
in the $(\textstyle{\frac12}K,q)$ plane where the integral on
$dK\,dq$ takes place,
depending on $k$ relative to $k_F$.

The above choice of coordinates facilitates tracking and control 
of dinucleon singularities.
Indeed, for constant $K$ the eigenenergy $\omega^*_{K}$ 
of an eventual dinucleon bound-state pole remains unchanged while 
the integration on $q$ takes place.
Additionally, considering the  spectral representation
of $g$ when bound states occur \cite{Arellano94}, 
its behavior for $\omega$ near the energy of the bound state becomes
\begin{equation}
\label{gomega}
g^{\alpha}_{K}(\omega)\sim 
\frac{vQ|\psi^{\alpha}_K\rangle\langle\psi^{\alpha}_K|Qv}
{\omega-\omega^*_K}\;,
\end{equation}
with $|\psi^{\alpha}_K\rangle$ the eigenfunction at the pole, $v$
the bare potential in the corresponding channel, and $Q$ the Pauli
blocking operator.
This analytic property allows us to safely state that the mass operator,
expressed as an integral over momenta of $g$ with isolated singularities, 
should be finite. 
The rationale behind this statement is that if simple poles occur, 
the integration over $q$ of the singular part of $g$ in 
eq.~(\ref{MassOperator}) reduces to
\begin{equation}
\label{integpole}
{\cal I} = 
\int_{q_{\textrm{min}}}^{q_{\textrm{max}}} \frac{2q\,dq}{k}\,
\frac{|\langle q|vQ|\psi^{\alpha}_K\rangle|^2}{E(K,q)-\omega^*_K} \;.
\end{equation}
Note that the numerator is bound.
Expanding the denominator to first order in $q$ around its zero,
which we assume inside the interval of integration, 
the contribution around the singularity can be reduced to the form
$\int_{-\delta}^{\delta}\textstyle{\frac{dx}{x}}$, which
becomes finite. 
Therefore, integrals over momenta of the $g$ matrix featuring 
dinucleon poles are not only finite but should be feasible.
In actual calculations these singularities are controlled with the 
regularization
\begin{equation}
\label{regularization}
g(\omega) \to g(\omega)\times 
\frac{(\Delta\omega)^2}{(\Delta\omega)^2+\eta^2}\;,
\end{equation}
where $\Delta\omega = \omega-\omega^*_K$, 
and $\eta$ is an infinitesimal. 
In this way contributions too close to the pole,
namely $\Delta\omega/\eta\to 0$, become attenuated.
On the other hand $g$ is practically unchanged whenever 
$|\Delta\omega|/\eta\gg 1$.
In the actual implementation of this technique we have used $\eta=0.1$~MeV.
These considerations in the handling of \SO and 
deuteron bound states render a much needed stability 
in the evaluation of the Fermi-motion integrals, which otherwise 
become uncontrollable.

Another aspect of particular relevance in the evaluation of $M$ in 
eq.~(\ref{MassOperator}) is the quadrature method to be used.
Two elements enter into consideration here. 
On the one hand it is the varying range of integration 
along $q$ for fixed $K$ 
(see fig.~\ref{FermiRegion});
on the other hand, it is the presence of bound states responsible for
steep variations of $g$.
These features combined make inadequate the use of Gaussian-type
quadratures designed for smooth integrands.

Aiming to a reliable precision in the evaluation of Fermi-motion integrals
we have resorted to an adaptive trapezoidal quadrature
considering gradually $2, 3, 5, \dots , (2^n+1)$ knots. 
The scheme, described in Appendix \ref{quadrature}, 
is first applied to the (innermost) $q$ quadrature and then 
to the integration over $K$.
The sequence is interrupted once the difference between two 
consecutive evaluations is bound to 2\%.
Given that in many cases the integrand is smooth and 
the spacing between consecutive knots diminishes exponentially, 
this criterion provides an overall accuracy in the sp fields 
of the order of 0.5\%.  
We have limited $n$ to eight, 
leading to 255 trapezoids over a given interval.
Self-consistency of the calculated $U(k)$ is achieved when the maximum 
fluctuations of $U(k_i)$ over three consecutive cycles 
do not exceed 0.04~MeV,
condition imposed at all $k_i$ where $U$ is evaluated.
In this work we have used $k\leq 5.5$~fm$^{-1}$.

\section{Results}
\label{results}
Self-consistent solutions for $U(k)$ were obtained in the range
$0< k_F\leq 1.75$~fm${}^{-1}$ using the Argonne $v_{18}$ bare internucleon
potential \cite{Wir95}.
We have included all \emph{NN} partial waves up to $J=7\hbar$, 
the total angular momentum.
This criterion was maintained at all densities in order to 
cross-check our findings near the saturation point, 
in addition to rule out any subtle variation of the calculated 
observables as caused by a discrete (sudden) change 
in the number of partial waves taken into account.
With these specifications and the numerical considerations described 
in the previous section, each cycle at which $U$ is calculated 
involves between $10^5$ and $10^6$ matrix inversion operations,
depending on the sharpness of the integrand. 

\subsection{Self-consistent solutions}
Usually the search for self-consistent solutions to the sp field 
by iterative feedback is stable enough to reach unique
solutions, even starting out with zero field.
We find this to be the case when the Fermi momentum is near
and above 0.35~fm${}^{-1}$.
The issue becomes more subtle at lower densities,
where the iterative process yields ill-behaved solutions.
This fact has led us to adopt different strategies for
$0<k_F < 0.35$~fm${}^{-1}$ and $k_F \geq 0.35$~fm${}^{-1}$.
In the second case we have found unique self-consistent solutions 
at Fermi momenta $k_F=0.35(0.05)1.75$~fm${}^{-1}$, following
the usual iterative procedures.

In contrast to this upper range, 
searches in the interval $0<k_F<0.35$~fm${}^{-1}$, 
required a specific strategy due to the fact that two different 
self-consistent solutions could be obtained for the same value of $k_F$.
These two genuine solutions meeting self-consistency 
will be referred to as \emph{coexisting solutions}.
To proceed systematically, a first class of solutions was obtained 
with ascending $k_F$, in steps of $\Delta k_F=0.01$~fm${}^{-1}$.
At each $k_F$, the self-consistent loop starts with an $U(k)$ 
borrowed from the converged solution at the previous $k_F$. 
In this way, starting at $k_F=0.01$~fm${}^{-1}$, we obtain a class of
self-consistent solutions characterized by a monotonically increasing
binding energy per nucleon,  $B/A$, as a function of $k_F$.
We denote this class of solutions as 
$U_{I}=\{U_{k_F}\mid 0\leq k_F\leq k_{\beta}\}$, 
with $k_{\beta}$ the maximum (critical) Fermi momentum at 
which self-consistency is achievable in ascending order.
To resolve $k_{\beta}$, self-consistent solutions were explored
with increasing $k_F$ in steps $\Delta k_F$ as small as 0.001~fm${}^{-1}$,
obtaining $k_{\beta}=0.285$~fm${}^{-1}$.
Beyond this point the iteration loops indefinitely, 
with fluctuations in $U(k)$ unable to settle below 0.04~MeV.
A more refined characterization of the solutions near and 
beyond $k_{\beta}$ may require a more specific approach.

A similar procedure was adopted for decreasing $k_F$, starting from the 
converged solution at $k_F=0.35$~fm${}^{-1}$ and going down 
in steps of 0.01~fm${}^{-1}$. 
As in the previous case, the continuity of $B/A$ is monitored and 
we seek the minimum critical value of $k_F$, denoted by $k_{\alpha}$, 
at which self-consistency is reachable.
This second class of solutions is denoted 
$U_{II}=\{U_{k_F}\mid k_F\geq k_{\alpha}\}$.
We obtain $k_{\alpha}=0.130$~fm${}^{-1}$, 
with a resolution of $0.01$~fm${}^{-1}$.

\begin{figure}
\begin{center}
\includegraphics[width=0.8\linewidth,clip=true] {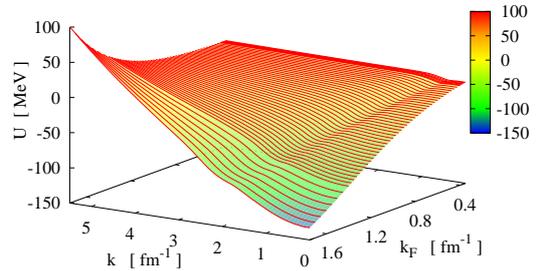}
\end{center}
\caption{{\protect
\label{UFieldsII}
 Surface plot for $U_{II}$ as a function of $k$ and $k_F$, in the case
 $k_F\geq 0.35$~fm${}^{-1}$. 
}}
\end{figure}

In fig.~\ref{UFieldsII} we show a surface plot of the
calculated self-consistent solutions $U_{II}$, 
as functions of $k$ and the Fermi momentum $k_F$, 
with $k_F\geq 0.35$~fm${}^{-1}$.
In general terms they follow the same trend as those reported
elsewhere \cite{Baldo00}, with $U(k)$ negative at $k=0$,
growing as $k$ increases.

\begin{figure}
\begin{center}
\includegraphics[width=0.8\linewidth,clip=true] {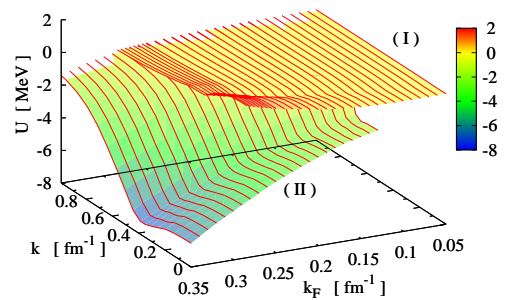}
\end{center}
\caption{{\protect
\label{UFieldsI}
 Surface plots for $U_I$ and $U_{II}$ as functions of $k$ and $k_F$, 
when $k_F\leq 0.35$~fm${}^{-1}$. 
}}
\end{figure}

In fig.~\ref{UFieldsI} we show surface plots for the self-consistent
fields $U(k)$ for solutions I and II in the range 
$0.05$~fm~${}^{-1}$$\leq k_F \leq 0.35$~fm${}^{-1}$. 
These solutions define distinct surfaces at densities 
corresponding to $k_{\alpha}\leq k_F\leq k_{\beta}$.
The upper sheet, representing $U_{I}$, features a moderate 
repulsion at $k=0$, with a shallow ditch in the vicinity $k\sim 4k_F/3$.
On the other hand 
the sheet for $U_{II}$ starts with negative mean fields at $k=0$, 
showing a moderate decrease up to $k\sim k_F$ and then followed by a 
steep increase to reach $U(k)\to 0$.
In both cases the slope of the self-consistent fields become negative
at the Fermi surface, feature responsible for nucleon 
effective masses greater than the bare nucleon mass, to be discussed ahead.

\begin{figure}
\begin{center}
\includegraphics[width=0.70\linewidth,clip=true] {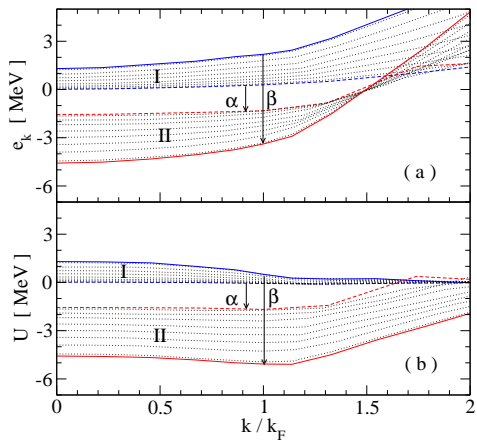}
\end{center}
\caption{{\protect
\label{ukbands}
Single-particle energies (a) and potentials (b) in the coexistence region
as functions of $k/k_F$.
Dashed and solid curves represent solutions for 
$k_F$ at $k_\alpha$ and $k_\beta$, respectively.
}}
\end{figure}

A closer view of the coexisting solutions is presented in 
fig.~\ref{ukbands}, where in panel (a) we plot sp energies as 
functions of the ratio $k/k_F$ in the coexistence region,
and in panel (b) the corresponding sp potentials.
Dashed and solid curves represent solutions for
$k_F=k_\alpha$, and $k_F=k_\beta$, respectively.
Dotted curves represent consecutive solutions for
$k_F=0.14(0.02)0.28$~fm${}^{-1}$.
Vertical arrows connect coexisting solutions at 
$k_\alpha$ and $k_\beta$, respectively.
The difference between Fermi energies of solutions I and II 
is 1.6~MeV at $k_\alpha$, and 5.6~MeV at $k_\beta$.
We also notice that none of the solutions for $U(k)$ (lower panel)
exhibit a parabolic behavior in the whole momentum range. 
As a matter of fact the slope of $U(k)$ for solution I exhibits a 
sudden increase from negative to near-zero at $k/k_F\approx 1.15$,
while for solution II the increase goes from near-zero to positive slope.
These features rule out the validity of the effective mass
approximation at low densities. 

\begin{figure}
\begin{center}
\includegraphics[width=0.7\linewidth,clip=true] {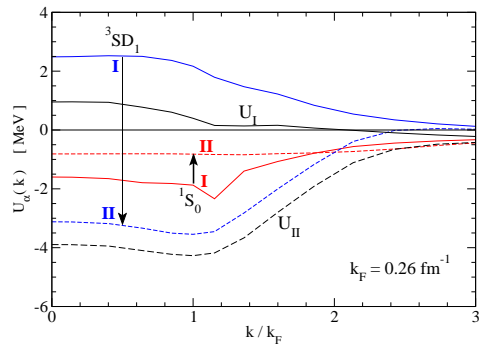}
\end{center}
\caption{{\protect
\label{ukpartial}
\SD and \SO partial contributions to the total sp 
potential $U(k)$ as functions of $k/k_F$ 
at $k_F=0.26$~fm${}^{-1}$. 
Solid and dashed curves represent phases I and II, respectively.
}}
\end{figure}

The disclosure of the two distinct phases I and II for the sp
spectrum constitutes an unexpected outcome of the calculations 
we have performed for nuclear matter in normal state.
To gain insight into possible causes behind this behavior,
in fig.~\ref{ukpartial} we plot the total 
sp potentials $U_I$ and $U_{II}$, together with the 
partial contributions from \SD and ${}^1\textrm{S}_0$ channels. 
Solid and dashed curves represent solutions in phase I and II,
respectively.
In this illustration we have selected $k_F=0.26$~fm${}^{-1}$,
a Fermi momentum reasonably close to $k_\beta$ but not at the edge 
of the overlap. 
The most prominent difference between phases I and II appears in
the deuteron channel, 
passing from a correlated Fermi gas (FG) behavior in phase I to 
an attractive (condensing) 
state in phase II, with an energy decrease of about 6~MeV near 
the Fermi surface ($k/k_F\approx 1$).
This is in contrast with the moderate increase, by about 1~MeV,
observed in the ${}^1\textrm{S}_0$ channel.
This feature points to the deuteron as the driving constituent responsible
of the occurrence of phase II beyond $k_\alpha$.

\begin{figure}
\begin{center}
\includegraphics[width=0.9\linewidth,clip=true] {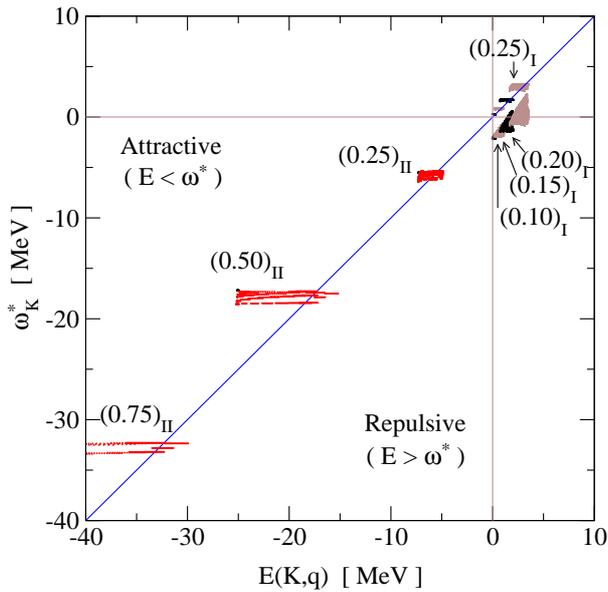}
\end{center}
\caption{{\protect
\label{wpoles}
Location of poles $\omega_K^*$ (if detected) 
and pair energy ($E=e_p+e_k$), upon
evaluation of $U(k)$ for $k\lesssim 1.2 k_F$.
The numbers in parenthesis represent $k_F$ (in fm$^{-1}$ units),
with the subscript referring to phase I or II. 
Variations of $(K,q)$ are in correspondence with domains shown 
in fig.~\ref{FermiRegion}.
}}
\end{figure}

The above comments are reinforced if we track \SO and \SD bound 
state occurrences while performing the Fermi motion 
integration [see eq.~(\ref{mass})] through the $(K,q)$ domains 
displayed in fig.~\ref{FermiRegion}.
In fig.~\ref{wpoles} each dot represents the energy of a detected pole, 
$\omega_K^*<\omega_{th}$, and the set of starting energies $E=e_p+e_k$, 
of the energy denominator in eq.~(\ref{integpole}) when integration on $q$ 
is performed. These dots are collected when $U(k)$ is evaluated for
discrete values of $k\lesssim 1.2 k_F$, typical momentum below which
poles are detected while performing Fermi-motion integrals.
Note that $E(K,q)=e_p+e_k$, when angular average is applied.
Dots appear separated as a result of the fact that (adaptive) 
quadratures are used for both $dq$ and $dK$ integrations.
In fig.~\ref{wpoles} a few selected values of $k_F$, 
indicated with numbers but omitting fm$^{-1}$ units, are chosen for clarity.
Clusters of dots in the form of `islands'
 displayed in the upper part of the panel, where $E(K,q)>0$,
correspond to phase I.
Out of these clusters, 
the ones leaning to the right hand-side with respect to the diagonal 
correspond to the \SD channel (subject to existence of a pole).
Similarly, the clusters in phase I being
crossed by the diagonal line correspond to the \SO channel.
From eq.~(\ref{integpole}) we infer positive (repulsive) contributions 
if $E(K,q)>\omega_K^*$, corresponding to the sector to the right hand-side
of the diagonal.
Conversely, attractive contributions stemming from dinucleons
are located in the left hand-side of the diagonal. 
In this `$E-\omega$' plot, the closer a dot is from the diagonal, 
the stronger is the associated contribution in the respective channel. 
The actual strength of the contribution 
will depend on the principal value of the pole.
In fig.~\ref{wpoles} the clusters shown for phase II correspond to
the \SD channel (very few or no poles are detected in the \SO channel),
which appear crossed by the diagonal but leaning toward the region $E<\omega^*$.
This indicates dominance of negative contributions in the
\SD channel, in contrast to phase I where clusters in same channel
lean toward positive contribution, \emph{i.e.} repulsive. 

The strategy we have adopted to obtain self-consistent solutions,
starting out iterative loops with a converged solution at a nearby $k_F$,
has proven essential to elucidate the coexisting solutions.
Actually if such a measure was not taken the calculated $U(k)$ might 
exhibit piecewise zigzagging instabilities, preventing any relaxation 
toward convergence.
In the case of solution I, successive solutions were obtained with
increasing $k_F$ up to $k_\beta$. 
When the solution at $k_F=k_\beta$ was used as starting guess 
for $k_F=k_\beta+\delta$, with $\delta$ some small increment, 
the requirement we have established for stable solutions could not be met
after sixty iterative loops.
However, if the increment is large enough the self-consistent
solution settles over that of $U_{II}$.
An analogous process is followed to obtain solution II, 
but generating solutions with decreasing $k_F$ up to $k_\alpha$.
Here again, the use of $U(k)$ at $k_\alpha$ to start iterations for
$k_F=k_\alpha-\delta$, would not relax if $\delta$ were 
small and positive, 
but it would settle over $U_I$ if $\delta$ were large enough.
In this respect, the procedure to obtain the solutions in the 
coexistence interval $[k_\alpha,k_\beta]$ resembles that of hysteresis.

The procedure described above was repeated all through the range 
$k_F \lesssim 0.3$~fm$^{-1}$, assuming valid the 
effective mass approximation while solving eqs. (\ref{bbg0}-\ref{usp})
subject to dinucleon bound state occurrences and self-consistency. 
We have been able to identify phase I, though over a more limited
range, when the quadratic approximation of $U(k)$ is imposed
over the range $0\le k\le 2k_F$. 
However, solutions for phase II become unstable, preventing actual 
self-consistency.
An specific study on merits and shortcomings of the
effective mass approximation goes beyond the scope of this work.

\subsubsection{Binding energy}
The fact that two different solutions for $U(k)$ satisfy self-consistency 
at a given Fermi momentum between $k_\alpha$ and $k_\beta$, 
implies two accessible sp states for nucleons in the medium.
If these were the only degrees of freedom of the system, 
then solution II would be the one accountable for the ground state.
However, pairing correlations beyond normal BHF 
would change the scenario,
an aspect that needs to be investigated within a more 
general framework \cite{Bozek2002}.
Having these shortcomings in mind, we have proceeded to evaluate the
binding energies per nucleon of the system considering phases 
I and II separately. This simplifying assumption would help us
to set bounds to the results.

\begin{figure}
\begin{center}
\includegraphics[width=0.8\linewidth,clip=true] {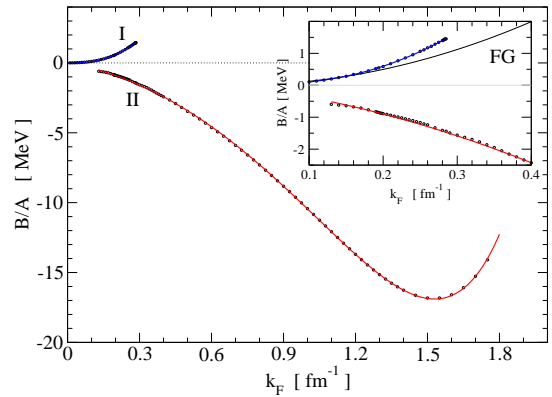}
\end{center}
\caption{{\protect
\label{boa}
Calculated $B/A$ 
for symmetric nuclear matter as a function of the Fermi momentum $k_F$.
Continuous curves represent the parametrization defined by eqs. (\ref{uu}).
Small circles denote actual results.
Inset shows a close up at low densities, including a free 
Fermi gas.
}}
\end{figure}

In fig.~\ref{boa} we plot the binding energy per nucleon for 
symmetric nuclear matter, $B_{I}/A$ and $B_{II}/A$,
as obtained from solutions $U_{I}$ and $U_{II}$, respectively.
Here each small dot denotes an actual calculation,
while the continuous curves represent suitable parametrizations.
We note that the two solutions exhibit distinct behaviors as functions
of $k_F$ in their respective domains, 
without intercepting each other in the range $[k_{\alpha},k_{\beta}]$.
While solution I resembles a correlated FG in metastable state, 
solution II represents a condensed medium featuring a minimum at the point
of nuclear saturation.
From the inset for low densities we also note that solution I
departs from an uncorrelated FG above $k_\alpha$, the Fermi momentum
where solution II begins.
This departure features a moderate repulsiveness relative to a free FG. 

To facilitate the analysis of solutions I and II
we have found simple parametrizations of the mean potential energy
\begin{equation}
\label{umean}
u(k_F) \equiv \frac{3}{k_F^3}\int_{0}^{k_F} k^2dk\,U(k)\;.
\end{equation}
Specifically
\begin{subequations}
\label{uu}
\begin{align}
u_I(x)    &= a_2(x/a)^2+a_3(x/a)^3+a_6(x/a)^6\;, \label{u1} \\
u_{II}(x) &= b_1(x/b)+b_2(x/b)^2+b_7(x/b)^7\;, \label{u2}
\end{align}
\end{subequations}
with $a=0.3$~fm$^{-1}$, and $b=1.53$~fm$^{-1}$.
These constructions lead to accurate representations of the 
calculated binding energy per nucleon
\begin{equation}
\label{epsil}
B_i/A = \frac{3}{10}\,\frac{k_F^2}{m}+\frac{1}{2}u_i(k_F)\;,
\end{equation}
with $i=I$ and $II$.
The coefficients $a_p$ and $b_p$, summarized in table \ref{coeffs},
were obtained from a least-square fitting procedures for the 
calculated $B_i/A$ values as a function of $k_F$.
The standard deviations associated with each set of coefficients 
are 0.01~MeV and 0.06~MeV, respectively.
In the case of solution $B_{II}/A$ 
we obtain a saturation energy of $-16.8$~MeV
at $k_F=1.53$~fm${}^{-1}$, with an incompressibility $K_\infty=213$~MeV.
\begin{table}
\centering
\caption{Best-fit coefficients for $u(k_F)$.}
\begin{tabular}{c c r | c c r }
\hline
\hline
& $p$ & \hspace{8pt}$a_p$ ( MeV )\hspace{8pt} & 
& $p$ & \hspace{8pt}$b_p$ ( MeV )\hspace{8pt} \\
\hline
& 2 & -1.25   \hspace{10pt}  && 1 &  -8.91   \hspace{10pt} \\
& 3 &  2.69   \hspace{10pt}  && 2 & -94.89   \hspace{10pt} \\
& 6 & -0.37   \hspace{10pt}  && 7 &  11.91   \hspace{10pt}  \\
\hline
\hline
\end{tabular}
\label{coeffs}
\end{table}

The coexistence of solutions I and II 
in the range $k_{\alpha}\leq k_F\leq k_{\beta}$ define 
three distinct regimes according to $k_F$,
namely:
\emph{(i)}
a diluted phase in the form of an interacting FG up 
to $k_{\alpha}$;
\emph{(ii)}
a mixed phase with coexisting sp states between $k_\alpha$ 
and $k_{\beta}$; and
\emph{(iii)}
a condensing phase above $k_\beta$.
The low-density behavior of $B/A$ in phase I, featuring a repulsive FG,
points to dominance of the kinetic contribution over its interaction term. 
Being this the case, if no condensation takes place, 
the gas behavior up to 
$k_F\sim 0.13$~fm$^{-1}$ ($\rho\sim 10^{-4}$~fm$^{-3}$), 
prevents homogeneous symmetric nuclear matter in normal state
from spontaneous collapse. 
As we shall see later, this repulsiveness gets greately attenuated 
if the sp momentum distribution accounts for pairing correlations.

A survey of available literature have resulted in scarce information
on zero-temperature and low-density behavior of symmetric nuclear matter
($k_F\lesssim 0.4$~fm$^{-1}$, or $\rho\lesssim 0.004$~fm$^{-3}$)
when the bare \emph{NN} interaction is used.
In ref. \cite{Baldo2004} for instance, the authors have made explicit 
the fact that a polynomial extrapolation is used for densities 
below 0.02~fm$^{-3}$ (\emph{i.e.} $k_F < 0.67$~fm$^{-1}$).
Extrapolations of similar nature, roughly from the same density, 
have been made in refs. \cite{Geramb83,Amos00} in the context 
of density-dependent effective interactions based on BHF model
and applied to nucleon-nucleus scattering.
In a recent publication \cite{Phat2011} aimed to the study of the
phase structure of symmetric nuclear matter in the extended 
Nambu-Jona-Lasinio model, 
fig.~1 for $B/A$ as a function of the density
suggests FG behavior at $\rho\lesssim 0.007$~fm$^{-3}$,
although no comments are made by the authors on this feature.

\subsubsection{Effective masses and dinucleon binding}
\begin{figure}
\begin{center}
\includegraphics[width=0.85\linewidth,clip=true] {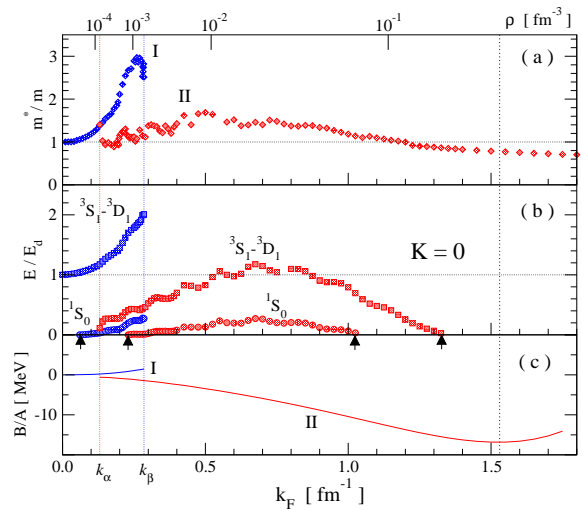}
\end{center}
\caption{{\protect
\label{mbe}
Effective mass relative to the bare mass (a);
dinucleon binding energies relative to $E_d=2.224$~MeV, for $K=0$ (b);
and binding energy per nucleon (c), as functions of $k_F$. 
The upper scale denotes nucleon density $\rho$.
The rightmost dotted line is drawn at saturation ($k_F=1.53$~fm${}^{-1}$).
Arrows in panel (b) denote loci where dinucleons get dissolved.
}}
\end{figure}

In this section we examine the density dependence of the effective 
mass and dinucleon binding energies associated to each phase in
normal state.
To this end we show in panel (a) of fig.~\ref{mbe} the total effective 
mass relative to the bare nucleon mass, $m^*/m$, as a function of 
the Fermi momentum.
Phase-I results are plotted up to $k_\beta$, while
results for phase II appear for $k_F\geq k_\alpha$.
In this case we evaluate
\begin{equation}
\label{masef}
\frac{m^*}{m} = 
\left [ 
1 + \frac{m}{k} \frac{\partial U(k)}{\partial k}
\right ]^{-1}\,,
\end{equation}
at $k=k_F$. 
In panel (b) we plot dinucleon binding energies in states
\SD (circles) and ${}^1\textrm{S}_0$ (squares), 
in units of deuteron binding in free space, $E_d=-2.224$~MeV.
The dinucleon binding energy $E$ is given by
\[
E=\omega-\omega_{\textrm{th}}\;,
\]
with $\omega$ the eigenenergy [see eq.~(\ref{det})] and 
$\omega_{\textrm{th}}$ the pp energy at the Fermi surface.
Here we restrict to cases where $K=0$, 
\emph{i.e.} pairs with no translational motion.
For reference purposes, in panel (c) we plot again the binding energy
per nucleon associated with solutions I and II.
The upper scale denotes density, $\rho=2k_F^3/3\pi^2$.
We first observe that $m^*/m$ for solution I grows monotonically
with increasing $k_F$, reaching a maximum $m^*/m\approx 3$, 
close to the upper edge of the coexistence interval ($k_\beta$).
In turn, $m^*/m$ for solution II starts out near unity at $k_\alpha$,
reaching a maximum $m^*/m\approx 1.8$, at $k_F=0.55$~fm${}^{-1}$.
Beyond $k_F=0.9$~fm${}^{-1}$, $m^*/m$ decreases monotonically, 
taking values from $0.88$ to $0.79$ for Fermi momenta 
from $1.3$ to $1.5$~fm${}^{-1}$, respectively, 
consistent with quotes found in the literature.
In the domain of coexistence
the two solutions yield dressed nucleons with distinct effective masses, 
reaching values as high as $m^*/m\approx 3$, for solution I and
$m^*/m\approx 1$, for solution II.

Panel (b) of fig.~\ref{mbe} shows qualitative similarities between 
deuteron and \SO binding energies,
although differences point to weaker binding in the case
of the ${}^1\textrm{S}_0$ channel.
Solution I for deuteron pairs shows increasing binding energy as the 
density grows up to $k_\beta$, reaching up to twice that in free space.
The same solution but for \SO state exhibits increasing binding up to
nearly $0.3\,E_d$, although no bound state is found in this channel 
for $k_F$ below $0.06$~fm${}^{-1}$ (Mott transition).
In the case of solution II, 
deuterons pairs are found for $k_F\geq k_\alpha$, 
but get dissolved beyond $k_F\approx 1.325$~fm${}^{-1}$.
In turn, \SO pairs appear over a narrower range in $k_F$: 
$0.23-1.05$~fm${}^{-1}$.
The strongest bindings are $E/E_d\approx 1.2$ and 0.3, for \SD and
\SO pairs, respectively. 
These maxima occur at $k_F\approx 0.7$~fm${}^{-1}$.
We also notice that the ${}^1\textrm{S}_0$ and \SD
binding energies tied with solution II display shallow extrema. 
These oscillating patterns are in contradistinction with the smooth 
behavior displayed over $k_F$ by conventional S and D solutions of
pairing gap equations in nuclear matter \cite{Vonder91}. 
We have found no interpretation as to why such pairing gaps and 
dinucleon binding energies, two closely related properties, 
differ in their patterns over $k_F$.

A simple argument of plausibility to admit effective masses greater 
than bare masses in a very diluted medium comes from considering 
two nucleons interacting \emph{via} an attractive square well.
If the depth of the potential is $V_0$ and its radius is $R$,
then the condition to hold at least one bound state is
$mV_0R^2 > {\pi^2}/{4}$.
In the context of two interacting neutrons in free space this 
condition should be regarded as barely missed, as inferred from 
the low-energy behavior of the phase shifts in 
the ${}^1\textrm{S}_0$ channel and scattering length $a_{nn}=-18.5$~fm.
If we assume unaffected the bare interaction by the medium, 
then \SO pairs become possible provided $m^*$ increases enough
so as to allow $m^*V_0R^2 > {\pi^2}/{4}$.
This trend is observed in panel (b) of fig.~\ref{mbe}, 
where the Fermi momenta at which \SO pairs get unbound are indicated 
with small black arrows on the horizontal axis. 
Here $k_F=0.06,\,0.23$ and $1.025$~fm${}^{-1}$,
corresponding to $m^*/m=1.06,\; 1.25$ and $1.15$, respectively.
For deuterons, instead, bound states occurrences appear 
in correspondence with $m^* \gtrsim 0.87m$.
In this case bound states dissolve near $k_F=1.325$~fm${}^{-1}$.
These features were cross-checked for solution I with the aid of 
a simple computer code applied to a square well potential 
where the trend of $m^*/m$ in panel (a) of fig.~\ref{mbe} 
can be reproduced by feeding in $E/E_d$ for the deuteron from panel (b).

\subsection{Dinucleons}
\label{dinucleons}
Singularities of the $g$ matrix in the real axis below the Fermi surface,
in the ${}^1\textrm{S}_0$ and \SD channels,
are unequivocal signs for \emph{in-medium} bound states. 
The former corresponds to a state with total isospin $T=1$, 
leading to three possible \emph{NN} configurations: 
proton-proton, proton-neutron and neutron-neutron.
The latter ($T=0$) represents a proton-neutron bound state, \emph{i.e.} 
deuteron.
In this section we explore some features of these
Cooper pair-like solutions \cite{Cooper56}, representing the formation
of single pairs in the presence of homogeneous nuclear matter
within normal state.
These bound states correspond to actual solutions of BHF equation.
Whether these objects become true \emph{in-medium} bound states requires
the inclusion of hole-hole propagation \cite{Typel10,Ropke09}.

\subsubsection{Bound states in normal BHF}
Let $E_b$ be the eigenenergy corresponding to a bound state
of two interacting nucleons in the nuclear medium.
Then it can easily be shown \cite{Dickh08} that 
\begin{equation}
\label{ieta}
\lim_{\eta\to 0} \,i\eta g(E_b+i\eta)= vQ|\psi\rangle\langle\psi|Qv\;,
\end{equation}
with $Q$ the Pauli blocking operator and $|\psi\rangle$ the corresponding 
eigenstate (dependence on ${\boldsymbol K}$ is implicit).
On the other hand, the wave equation associated with eq.~(\ref{bbg0}) 
when $\omega$ matches the eigenenergy $E_b$ reads 
\[
(\hat h_1 +\hat h_2 + QvQ)| \psi \rangle = E_b |\psi \rangle\;.
\]
In free space, 
where $(\hat h_1+\hat h_2)$ becomes the  kinetic energy of the 
two nucleons and $Q$ the identity,
this equation reduces to the Schr\"odinger equation in stationary state.
After projecting in momentum space ($\langle{\boldsymbol q}|$) we
obtain for the wavefunction
\begin{equation}
\label{psi}
\langle{\boldsymbol q}|\psi\rangle=
\frac{\langle{\boldsymbol q}|QvQ|\psi\rangle}{E_b-e_{+} - e_{-}}\;,
\end{equation}
where $e_{\pm}= e(k_{\pm})$, 
consistent with the notation in eq.~(\ref{kpm}).
In the case of pairs without translational motion,
where $K=0$, the numerator reduces to
\begin{equation}
\label{qvq}
\langle{\boldsymbol q}|QvQ|\psi\rangle=
\Theta(q-k_F)
\langle{\boldsymbol q}|vQ|\psi\rangle \;.
\end{equation}
Thus, the wavefunction can be obtained 
after $\langle{\boldsymbol q}|vQ|\psi\rangle$ 
is extracted from eq.~(\ref{ieta}), 
a procedure we have achieved numerically solving the BHF 
equation for small $\eta$ and extrapolating to $\eta\to 0$.
The solution in momentum space is then Fourier transformed to get its 
coordinate space representation.

\subsubsection{Eigenfunctions}

As mentioned earlier, wavefunctions are calculated in momentum space.
If $\psi(q)$ denotes the wavefunction for a bound state of orbital
angular momentum $L$, its coordinate space representation is given by
\begin{equation}
\label{fourierL}
\Psi(r) = \sqrt{\frac{2}{\pi}} 
\int_{\bar q}^\infty q^2\,dq\,j_L(qr)\psi(q)\;.
\end{equation}
The lower integration bound $\bar q$ corresponds to that allowed
by Pauli blocking. In what follows we shall consider
pairs with their center of mass at rest 
$(K=0)$, so that $\bar q=k_F$.

To achieve reliable precision in the evaluation
of $\Psi(r)$ we found it crucial to control the behavior
of $\psi$ near the Fermi surface. 
To this end we constructed auxiliary functions 
\begin{equation}
\label{psi0}
\psi_0(q) = A q^L\exp[-R(q-k_F)]\,\Theta(q-k_F)\;,
\end{equation}
with $\Theta$ the Heaviside step function to suppress momentum
components below the Fermi surface. 
Parameters $A$ and $R$ are adjusted to match the exact $\psi(q)$ just 
above the surface, \emph{i.e.}
$q\to k_F^+$. With this construction we evaluate
\begin{equation}
\label{fourier}
\Psi(r)=\sqrt{\frac{2}{\pi}}
\int_{k_F}^\infty q^2\,dq\,j_L(qr)(\psi-\psi_0)
+ \Psi_0(r)\;,
\end{equation}
where
\begin{equation}
\label{fourier0r}
\Psi_0(r)=\sqrt{\frac{2}{\pi}}
\int_{k_F}^\infty q^2\,dq\,j_L(qr)\psi_0(q)\;.
\end{equation}
For the latter we proceed analytically, where specific results
are presented in Appendix \ref{exactfourier}.
Integrals are evaluated over a finite intervals
in $r$ and $q$, respectively. 
These intervals are chosen so that
\[
\int_0^{R_{\textrm{max}}} |\Psi(r)| \,r^2 dr =
\int_{k_F}^{q_{\textrm{max}}} |\psi(q)|^2\, q^2 dq\;,
\]
is met with reasonable precision.
Typically $q_{\textrm{max}}\approx 15$~fm${}^{-1}$, 
whereas $R_{\textrm{max}}$ could go
as high as $1500$~fm.

\begin{figure}
\begin{center}
\includegraphics[width=0.9\linewidth,clip=true] {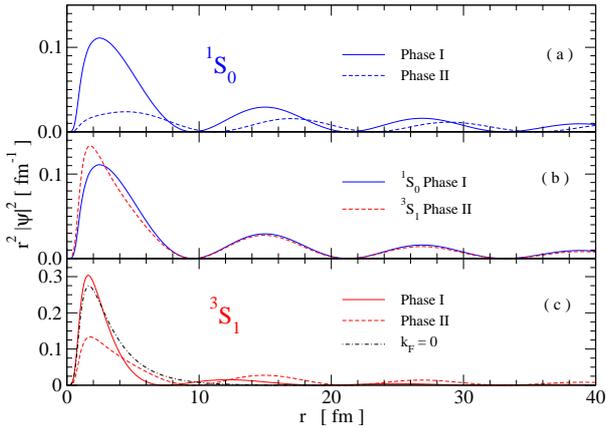}
\end{center}
\caption{{\protect
\label{waves}
Radial probability density, $r^2|\Psi(r)|^2$,
for \emph{in-medium} 
S-wave dinucleon bound state solutions as functions of 
the relative distance $r$.
Solid (dashed) curves denote solutions for phase I (II) in the
case of $k_F=0.25$~fm${}^{-1}$.
Panels (a) and (c) show results for channels 
${}^1\textrm{S}_0$ and ${}^3\textrm{S}_1$, respectively.
Panel (b) displays ${}^1\textrm{S}_0$ and \SD bound-state waves 
in phase I and II, respectively.
The dash-dotted curve in panel (c) represents the deuteron in free space.
}}
\end{figure}

In fig.~\ref{waves} we plot radial probability densities, 
$r^2|\Psi(r)|^2$, 
for S-wave bound states as functions of the relative distance $r$.
These solutions correspond to $k_F=0.25$~fm${}^{-1}$, 
Fermi momentum where the domains of phases I and II overlap.
Panels (a) shows results for the ${}^1\textrm{S}_0$ channel 
in phase I (solid curve) and phase II (dashed curve).
The same notation is used in panel (c) for the ${}^3\textrm{S}_1$ channel,
where we also include the S-wave deuteron in free space ($k_F=0$).
Panel (b) displays ${}^1\textrm{S}_0$ and 
${}^3\textrm{S}_1$ bound-state waves in phase I and II, respectively.

We observe that both ${}^1\textrm{S}_0$ and ${}^3\textrm{S}_1$ probability
densities represent confined systems, exhibiting the same oscillatory
behavior of Cooper pair solutions reported in 
refs. \cite{Baldo95,Matsuo06}.
In the case of the ${}^1\textrm{S}_0$ bound state shown in panel (a)
we notice an outward shift of phase II solution 
relative to that in phase I.
This indicates an enlargement of \SO pairs 
in phase II relative to those in phase I, to be discussed later.
The same feature is observed for deuterons in panel (c), 
where the free-space solution (solid black curve) is included. 
In this case we observe a similarity between the probability density of 
deuterons in free space with that of \SO pairs in phase I.
This suggests that deuterons maintain their size as long as they
remain in phase I. 
These spatial aspects of \emph{in-medium} dinucleon solutions shall 
further be discussed in the following section.

\begin{figure}
\begin{center}
\includegraphics[width=\linewidth,clip=true]{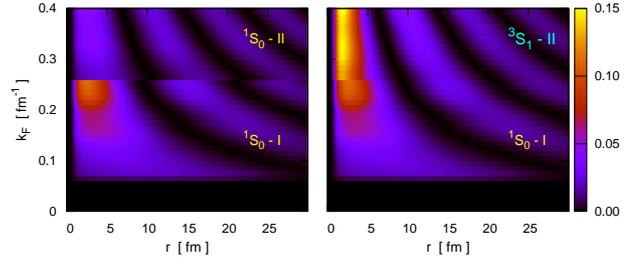}
\end{center}
\caption{{\protect
\label{merging}
\emph{In-medium} S-wave bound state radial 
probabilities $r^2|\Psi(r)|^2$ as functions of $r$ and $k_F$.
The left hand-side panel displays ${}^1\textrm{S}_0$ solutions 
(phase I, with $k_F\leq 0.26$~fm${}^{-1}$, 
and phase II with $k_F\geq 0.26$~fm${}^{-1}$). 
The right hand-side panel is the same as the left one, but with the
${}^1\textrm{S}_0$ solution in phase II being replaced by the
corresponding ${}^3\textrm{S}_1$ solution. 
}}
\end{figure}

In fig.~\ref{merging} we show contour plots
for S-wave radial probabilities $r^2|\Psi(r)|^2$ as 
functions of the relative distance $r$ and Fermi momentum $k_F$.
The left hand-side panel shows the ${}^1\textrm{S}_0$ solutions 
in phase I, 
with $0\leq k_F\leq 0.26$~fm${}^{-1}$, together with those in
phase II for $k_F\geq 0.26$~fm${}^{-1}$.
In the right hand-side panel the ${}^1\textrm{S}_0$ solution in phase II
has been replaced by the corresponding ${}^3\textrm{S}_1$ solution.
It is interesting to note how the two solutions appear in phase 
near the boundary $k_F\sim 0.26$~fm${}^{-1}$.

\subsubsection{Spatial properties}
\label{spatial}

Cooper-like wavefunctions appear to have unique spatial properties,
especially regarding their large size relative to the mean internucleon
distance in the nuclear medium. This feature is already suggested in
figs. \ref{waves} and \ref{merging}, where the bound state
wavefunctions are observed to decay slowly.
Actually, these wavefunctions behave as $\sim\cos(k_Fr)/r^2$ for 
large radii, 
feature which poses convergence difficulties in the actual evaluation
of mean radii $r_m\equiv\langle r\rangle$ or root-mean-square (rms) radii
$\langle r^2\rangle^{1/2}$.
In momentum space, in turn, we cannot make straightforward use of the 
customary identity \cite{Sun2010} for the mean-square-radius 
\[
\langle r^2 \rangle=
\frac{\int |\Psi_{\textrm{pair}}(r)|^2r^4dr}
{\int |\Psi_{\textrm{pair}}(r)|^2r^2dr}=
\frac{\int_0^\infty (\frac{\partial\psi}{\partial q})^2q^2dq}
{\int_0^\infty \psi^2q^2dq}\;,
\]
due to the discontinuity of the wavefunction $\psi(q)$ at the 
Fermi surface.
Actually, the above identity is valid as long as
the wavefunction is bound in coordinate space, while in momentum space
derivatives are continuous at all momenta.
Such is not the case for $\psi(q)$, exhibiting  
a cut-off below $q=\bar q$, as a result of the sharpness of the
sp momentum distribution.
This cut off occurs at $k_F$ for pairs with no translational motion ($K=0$).

To overcome the convergence difficulties posed in coordinate 
space evaluations, we propose to consider the Laplace transform 
of the density $r^2|\Psi(r)|^2\equiv r^2\rho$.
This transform reads
\begin{equation}
F(s)={\cal L}\{r^2\rho(r)\} = 
\int_{0}^\infty e^{-sr}\, |\Psi(r)|^2 r^2 dr\;.
\end{equation}
Clearly, $F(0)={\cal N}$, the volume integral of the density.
The small-$s$ behavior yields
\[
F(s) = {\cal N} \, \left ( 1 - \langle r \rangle\,s + 
\textstyle{\frac{1}{2}}\langle r^2\rangle \,s^2 - \cdots \right ) \;,
\]
from which we construct the alternative function $f(s)$ given by
\begin{equation}
\label{laplace}
f(s)\equiv \frac{1}{s}\left [ 1 - \frac{F(s)}{\cal N} \right ] =
\langle r \rangle - 
\textstyle{\frac{1}{2}}\langle r^2\rangle \,s + \cdots\;.
\end{equation}
Since ${\cal N}$ is finite,
$F(s)$ becomes always defined due to the exponential in the integrand.
Thus, from an analysis of $f(s)$ at small $s$ we 
can extract the mean radii $\langle r \rangle$
and the mean-square-radii $\langle r^2 \rangle$.
A linear regression of eq.~(\ref{laplace}) for $f$ 
provides us with accurate fits, 
with correlations equal to unity out to six significant figures.

\begin{figure}
\begin{center}
\includegraphics[width=0.90\linewidth,clip=true]{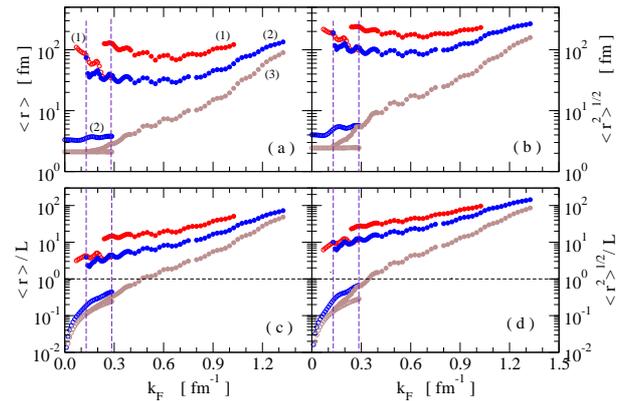}
\end{center}
\caption{{\protect
\label{xover}
Mean radii for \emph{in-medium} 
${}^1\textrm{S}_0$, 
${}^3\textrm{S}_1$ and
${}^3\textrm{D}_1$ bound states,
labeled (1), (2) and (3), respectively,
as functions of $k_F$.
Panel (a) and (b) show mean and rms radii, respectively.
Panels (c) and (d) show the corresponding quantities but relative to
$\rho^{-1/3}$.
}}
\end{figure}

In fig.~\ref{xover} we show results for the calculated mean radii 
$\langle r \rangle$ and rms radii $\langle r^2 \rangle^{1/2}$
as functions of the Fermi momentum $k_F$. 
Solutions for states 
${}^1\textrm{S}_0$, ${}^3\textrm{S}_1$ and ${}^3\textrm{D}_1$ are
labeled as (1), (2) and (3), respectively.
Empty and filled symbols denote results for phases I and II, respectively.
Upper panels show $\langle r\rangle$ (a) 
and $\langle r^2 \rangle^{1/2}$ (b) in fm units.
Lower panels show results for the ratios $\langle r\rangle/L$ (c) and 
$\langle r^2 \rangle^{1/2}/L$ (d),
with $L=\rho^{-1/3}$, the mean internucleon separation at density $\rho$.
We note that both mean radii and rms radii for 
\SO pairs (1) are larger than those for deuterons (2 and 3).
What is also evident is that S-wave bound states experience a sudden 
increase of size in the transit from phase I to phase II. 
The rate of the change is more pronounced for deuterons 
than for \SO pairs.
Additionally, observing panels (c) and (d) we note that
deuteron bound states in phase I are much smaller than the 
internucleon separation $L$,
while in phase II they remain bound with sizes several times 
the internucleon separation. 

Another feature worth noting from panel (a) is that the size of deuterons
in phase I remain roughly constant as a function of the density 
($k_F\leq\,0.285$~fm${}^{-1}$), with sizes similar to that in free space.
In phase II, instead, they become fairly large objects with mean radii 
above 30~fm.
In this respect deuterons change from very compact bosons in phase I
to very extended ones in phase II. 
This is in contrast to \SO pairs, where their sizes 
outpass the internucleon separation in both phases.

\subsection{BCS pairing}
The occurrence of dinucleon bound states in nuclear medium is
closely related with nuclear pairing phenomena, mechanism
responsible for the formation of Cooper pairs and emergence of 
superfluid and superconducting states of matter 
\cite{Broglia13,Sedrakian06}.
Hence, it is of interest to explore some immediate implications 
of phases I and II for the sp potentials in the context of pairing 
gap equations.
To this end we have solved the
Bardeen-Cooper-Schrieffer (BCS) gap equations at zero temperature.
These gap equations in triplet coupled states lead to coupled
equations displaying an explicit angular dependence of the kernel, 
problem which has been addressed in ref. \cite{Takat93} in the
context of pure neutron matter.
However, an important simplification to the problem is obtained
when the anisotropic kernel is angle averaged \cite{Dean03,Lombardo99}. 
Within this approximation the gap functions for states of orbital 
angular momentum $L$ take the form
\begin{equation}
\label{bcs}
\Delta_L(k) = -\frac{2}{\pi}\int_0^\infty k'^2\,dk' 
\sum_{L'} 
v_{LL'}(k,k')\,\frac{\Delta_{L'}(k')}{2E(k')}\;,
\end{equation}
where the quasiparticle energy reads
\begin{align}
\label{eqp}
E(k)^2 &= (e_k-\mu)^2 +\Delta(k)^2,\\
\Delta(k)^2 &= \sum_L\Delta_L(k)^2\;.
\end{align}
Here $\mu$ is the chemical potential and $e_k=k^2/2m+U(k)$, 
corresponds to the BHF sp spectrum.
In this case solutions I and II for $U(k)$ are treated independently,
an issue that needs to be reexamined in more realistic calculations.
The matrix elements $v_{LL'}(k,k')$ of the bare interaction
in a channel of total spin $S$ and isospin $T$ 
are given by
\begin{equation}
\label{qbare}
v_{LL'}(k,k') = 
i^{L-L'}
\int_0^\infty r^2\,dr\,
j_{L}(kr)v_{LL'}(r)\,j_{L'}(k'r)\;.
\end{equation}
The corresponding normal ($n$) and anomalous ($\kappa_L$) 
density distributions are given by
\begin{equation}
\label{normal}
n(k)=\frac12 \left[1-\frac{e_k-\mu}{E(k)}\right], \quad 
\kappa_L(k)=\frac{\Delta_L(k)}{2E(k)}\;, 
\end{equation}
respectively.
Self-consistency for the chemical potential is imposed through
\begin{equation}
\label{chemical}
\rho = 4\int \frac{d^3k}{(2\pi)^3} n(k)\;,
\end{equation}
with $\rho$ the nucleon density.

We have developed computer codes to solve BCS 
gap equations [eqs. (\ref{bcs}-\ref{chemical})] following
the method introduced by Baldo \emph{et al.} \cite{Baldo90}.
The approach relies on the introduction of a reduced interaction
which disconnects the low and high momentum components of the
interaction but reproduces exactly the gap function in the
low momentum regime.

\begin{figure}
\begin{center}
\includegraphics[width=0.8\linewidth,clip=true]{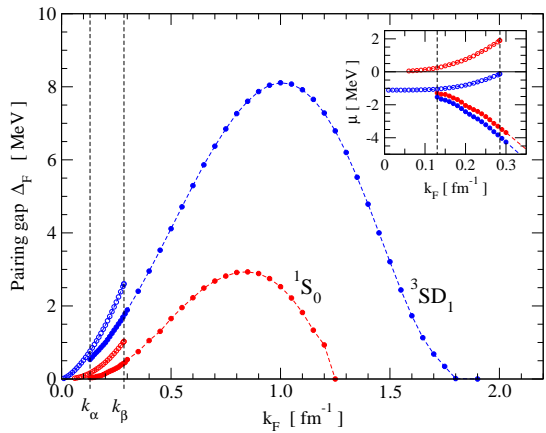}
\end{center}
\caption{{\protect
\label{gapsu}
Energy gap $\Delta_F$ as function of $k_F$
in symmetric nuclear matter for channels ${}^1\textrm{S}_0$ and \SD. 
Empty and filled symbols denote results for phases I and II, respectively.
Inset shows the chemical potential $\mu$ as function of $k_F$.
}}
\end{figure}

In fig.~\ref{gapsu} we show results for pairing gaps 
$\Delta_F=\Delta(k_F)$, in symmetric nuclear matter for 
${}^1\textrm{S}_0$ and \SD states.
Empty and filled circles denote the use of BHF potentials
in phase I and II, respectively. 
The inset shows the corresponding chemical potentials at low densities. 
Gap calculations in the overlapping domain of phases I and II
have been performed in a simplified way, namely, 
pairing in phase I do not take into 
account phase II and \emph{vice versa}. 
This simplification is intended to illustrate the most immediate 
consequences of the coexisting sp solutions. 

Overall we observe that pairing gaps as function of the Fermi momentum,
when $k_F\gtrsim 0.3$~fm${}^{-1}$, 
behave similarly to those published elsewhere \cite{Broglia13}.
The ${}^1\textrm{S}_0$ state in symmetric nuclear matter
yields non-vanishing pairing gap up to $k_F\approx 1.25\,\textrm{fm}^{-1}$,
with a peak $\Delta_F=2.93$~MeV, at $k_F=0.85\,\textrm{fm}^{-1}$.
Similarly, pairing in the \SD channel occurs
up to $k_F\approx 1.8\,\textrm{fm}^{-1}$, 
with a peak $\Delta_F=8.1$~MeV, at $k_F=1.0\,\textrm{fm}^{-1}$.
For $k_F\lesssim 0.3$~fm${}^{-1}$, however, the coexisting phases for the
sp potentials yield different energy gaps.
These features contrast with the continuous low-density solutions for 
symmetric nuclear matter based on Gogny's interaction \cite{Matsuo06}.
It is also worth noting the correspondence between the Fermi momenta
at which BCS pairing disappears and those where dinucleons in normal state
within BHF get dissolved [panel (b) of fig.~(\ref{mbe})].
What also becomes clear is that the maximum pairing gap in the 
\SO channel is about 1/3 that in the \SD state.
Considering that \emph{the} gap is given \cite{Lombardo99} by
$\Delta(k)^2 = \Delta_{^1S_0}(k)^2 + \Delta_{^3SD_1}(k)^2 +\cdots$,
this results in the actual suppression of \SO pairing 
in symmetric nuclear matter when the coupled \SO--\SD gap equations 
are solved.
From now on we shall only consider pairing in the \SD channel.

Regarding the low density behavior of the chemical potential shown
in the inset of  fig.~\ref{gapsu}, 
we note that for the deuteron channel $\mu\to-1.1$~MeV,
close to half the binding energy of the deuteron in free space. 
This is an expected result as it is well known that gap equations 
at low densities reduce to the Schr\"odinger equation in free space,
where the anomalous density distribution $\kappa$ takes the 
place of the wavefunction.

\begin{figure}
\begin{center}
\includegraphics[width=0.85\linewidth,clip=true]{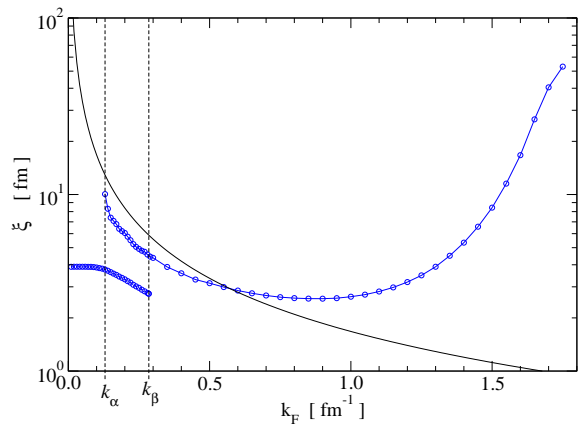}
\end{center}
\caption{{\protect
\label{xi}
BCS coherence length in the \SD channel as a function of $k_F$ 
in symmetric nuclear matter.
The solid curve represents the mean internucleon separation $\rho^{-1/3}$. 
}}
\end{figure}

In BCS theory the coherence length $\xi$ sets a length scale 
for the distance between the two constituents forming Cooper pairs.
In terms of the anomalous density distribution, this is obtained from
\[
\xi^2 = 
\frac{\int_0^\infty (\frac{\partial\kappa}{\partial k})^2k^2dk}
{\int_0^\infty \kappa^2k^2dk}\;.
\]
The evaluation of $\xi$ follows the procedure introduced 
in the previous section based on Laplace transforms, this time 
using the anomalous density. 
In fig.~\ref{xi} we plot
the coherence length as function of $k_F$ for the \SD channel.
The solid curve represents the internucleon separation $\rho^{-1/3}$.
Note that the coherence length in phase I amounts between 0.4
and 0.6 times that in phase II.
Therefore Cooper pairs in phase I are more compact than those in phase II,
in correspondence with dinucleon radii.
Additionally, $\xi$ in phase II outpasses the internucleon separation
at $k_F\approx 5.5$~fm$^{-1}$, indicating that Cooper pairs 
extend beyond the mean separation between nucleons. 
Conversely, at low densities these objects become compact in the sense that
their size is significantly smaller than the internucleon separation.

\begin{figure}
\begin{center}
\includegraphics[width=0.7\linewidth,clip=true]{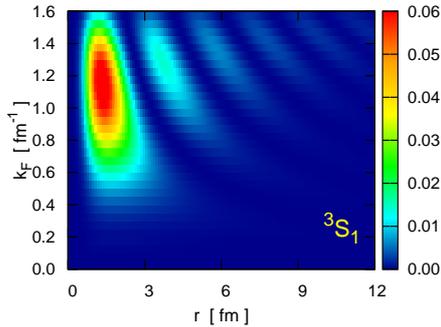}
\end{center}
\caption{{\protect
\label{anomal}
Contour plot for the 
${}^3\textrm{S}_1$ 
anomalous probability amplitude $r^2|\Psi(r)|^2$ (in fm$^{-1}$ units)
as function of the relative distance $r$ and Fermi momentum $k_F$. 
}}
\end{figure}

In fig.~\ref{anomal} we show a surface plot of the 
${}^3\textrm{S}_1$ anomalous probability density
$r^2|\Psi(r)|^2$ calculated from the anomalous density distribution,
as functions of the relative distance $r$ and Fermi momentum $k_F$.
These wavefunctions have been calculated using the sp potentials in 
phases I and II.
We observe how the wavefunction spreads over greater extent of space
as density increases, consistent with the behavior of the correlation
length $\xi$ shown in fig.~\ref{xi}.
Additionally, the Fermi momenta at which these functions peak appear
in correspondence with the pairing gaps shown in fig.~\ref{gapsu}.

\begin{figure}
\begin{center}
\includegraphics[width=0.85\linewidth,clip=true]{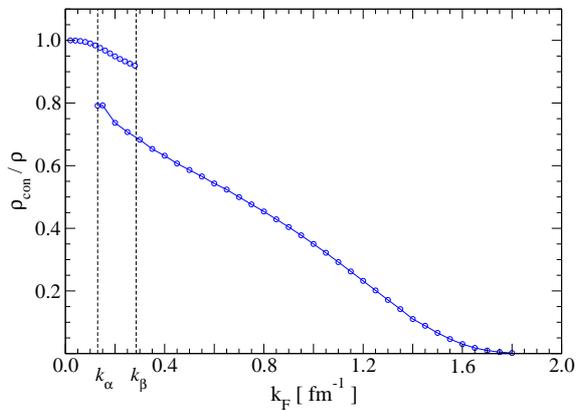}
\end{center}
\caption{{\protect
\label{condensate}
Condensate fraction in \SD channel as function of the Fermi momentum, 
considering sp spectra in phases I and II separately.
}}
\end{figure}

Having solved gap equations for phases I and II separately, 
we can now evaluate the condensate fraction in the \SD channel.
In terms of the normal and anomalous density distributions the condensate
fraction of paired nucleons is given by \cite{Salasnich11}
\begin{equation}
\label{condens}
\frac{\rho_{\textrm{con}}}{\rho}=
\frac{\int_0^\infty \kappa^2(k)\,k^2 dk}
     {\int_0^\infty n(k)\,k^2 dk}\;.
\end{equation}
In fig.~\ref{condensate} we plot the deuteron condensate fraction 
$\rho_{\textrm{con}}/\rho$ as a function of the Fermi momentum.
We observe nearly total deuteron condensate (above 90\%) at low densities, 
in phase I, accounting for Bose-Einstein condensation. 
The condensate fraction in phase II decreases 
monotonically from $0.8$ at $k_F=k_\alpha$, 
vanishing at $k_F\sim 1.8$~fm${}^{-1}$.

\subsection{Superfluidity}
\label{superfluid}

As mentioned earlier, 
the results discussed above help us to visualize the trend of nuclear
pairing when each phase is considered separately. 
Additionally, the sp energies have been obtained within the BHF 
framework in normal state. 
The inclusion of superfluidity alters the momentum distribution
and subsequently the sp fields \cite{Lombardo99}, leaving open
the issue on coexistence of the two phases found for the normal
system.
A consistent treatment of the problem would require considerably
more efforts. 
Nonetheless, it is possible at the lowest order to estimate some
immediate implications of superfluidity on the system along the
line discussed in ref.~\cite{Lombardo99}.

\begin{figure}
\begin{center}
\includegraphics[width=0.85\linewidth,clip=true]{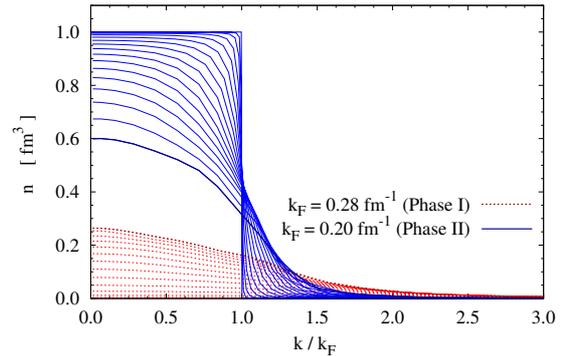}
\end{center}
\caption{{\protect
\label{nk}
Normal density as function of $k/k_F$ for phase I (dashed curves) and
phase II (solid curves).
}}
\end{figure}

In fig. \ref{nk} we plot the normal density $n(k)$ at various $k_F$ 
as function of the ratio $k/k_F$.
Phase-I results (dashed curves) are shown up to $k_F=0.28$~fm$^{-1}$,
in steps of 0.02~fm$^{-1}$.
Phase-II results are shown from $k_F=0.2$ up to 1.8~fm$^{-1}$
in steps of 0.1~fm$^{-1}$.
Note that the momentum distribution at high $k_F$ resembles very 
closely that of a normal Fermi distribution, consistent with
the suppression of BCS pairing.
As the density diminishes the momentum distribution spreads over
$k_F$ and gets depleted below $k_F$, in correspondence with
the onset of superfluidity.
This feature gets more pronounced in phase I.

To gauge the implications in the change of the momentum
distribution when pairing is included, we have evaluated the
energy per nucleon of the system including the condensation
energy \cite{Lombardo99,Fetter}, namely
\begin{equation}
\label{boabcs}
\frac{B}{A} = \frac{1}{\rho} 
\sum_{k}
\left\{ 
4n(k) 
\left [
\frac{k^2}{2m} + \frac{1}{2} U(k)
\right ]
-2\frac{\Delta^2(k)}{2E(k)}
\right \}\;.
\end{equation}
The term from this expression which contains the sp fields $U(k)$ is
denoted as $U_{BHF}$, 
\begin{equation}
\label{ubhf}
U_{BHF} = \frac{2}{\rho} 
\sum_{k} n(k)\, U(k) \;,
\end{equation}
whereas the last term represents the condensation energy and is 
denoted by $U_{BCS}$,
\begin{equation}
\label{ubcs}
U_{BCS} = -\frac{1}{\rho} 
\sum_{k} \frac{\Delta^2(k)}{E(k)}\;.
\end{equation}
For the evaluation of $B/A$ above we use $n(k)$ as obtained from
eq.~(\ref{normal}) based on sp energies in phase I and II, while
the sp potential corresponds to that of normal BHF.
Therefore it is not self-consistent in the sense that the evaluation 
of the sp potentials, $U(k)$, does not take into account the modification 
of the sp momentum distribution.

\begin{figure}
\begin{center}
\includegraphics[width=0.9\linewidth,clip=true]{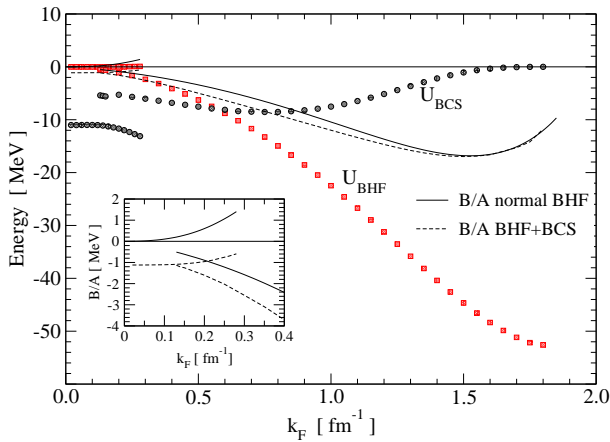}
\end{center}
\caption{{\protect
\label{boa_bhfbcs}
Calculated $B/A$ as obtained
from normal BHF (solid curves) and including condensation
energy (dashed curves).
The inset shows a close-up at low densities, including 
solutions in phase I and II.
Circles denote the condensation energy ($U_{BCS}$) 
and squares the BHF mean field contribution ($U_{BHF}$).
}}
\end{figure}

In fig. \ref{boa_bhfbcs} we present the calculated $B/A$ as obtained
from normal BHF (solid curves) and that using eq.~(\ref{boabcs}),
labeled `BHF+BCS'. 
The inset shows a close-up at low densities, where the solutions from the 
two phases are displayed.
Circles denote the condensation energy ($U_{BCS}$) 
and squares the BHF mean field contribution ($U_{BHF}$).
Note that the condensate energy is weak at high densities
but increases considerably at low densities, 
becoming the dominant (attractive) contribution in $B/A$.
The inclusion of pairing in phase I yields $B/A\approx -1.1$~MeV,
as observed from the inset (dashed curve in phase I). 
This feature is consistent with the fact that
nearly total condensation occurs at low densities.
What is also interesting to note is that the condensation energy 
intercepts $U_{BHF}$ at $k_F\approx 0.55$~fm$^{-1}$,
about the same Fermi momentum at which the correlation 
length $\xi$ outpasses the mean internucleon 
separation $\rho^{-1/3}$ (see fig.~\ref{xi}).
Another result which emerges when pairing correlations
are included is a slight decrease of the saturation density, 
up to about $k_F\approx 1.5$~fm$^{-1}$, 
without altering the saturation energy.

From the results discussed above it becomes clear that the 
disentanglement of BHF and superfluidity energies is meaningful
at normal densities, \emph{i.e.} near saturation,
in contrast to low densities where the condensation energy
amounts to a significant contribution to the total energy.
This result leaves open the question on whether coexisting 
solutions persist if a more complete treatment of superfluidity
is made.

\section{Summary and outlook}
\label{fin}
Within the BHF approximation in BBG theory for symmetric nuclear matter
at zero temperature,
we have investigated the role of dinucleon structures over a wide
range of densities 
($10^{-7}\,\textrm{fm}^{-3}\lesssim\rho\lesssim 0.36\,\textrm{fm}^{-3}$),
with emphasis placed on the low-density regime.
To this end we have calculated self-consistent sp potentials at Fermi 
momenta up to 1.75~fm${}^{-1}$ using the continuous choice, 
restricting the system to normal state.
The study is based on the Argonne $v_{18}$ bare \emph{NN} potential
considering all partial waves up to $J=7\hbar$, without three-body forces. 
The actual momentum dependence of the sp potential resulting from the 
evaluation of the mass operator has been retained.
An explicit treatment was given to the occurrence of 
dinucleon bound states in the \SO and \SD channels
during the evaluation of the mass operator.

As a result, two distinct families of solutions for the sp self-consistent 
potentials have been disclosed.
The first class, $U_I$, defines a surface in the $(k,k_F)$ plane
at Fermi momenta below $k_\beta=0.285$~fm${}^{-1}$.
The second class, $U_{II}$, defines a surface at 
Fermi momenta above $k_\alpha=0.130$~fm${}^{-1}$.
In the range $[k_\alpha,k_\beta]$ both solutions satisfy 
the self-consistency requirement, representing coexisting solutions.
Solution I behaves as a correlated FG with slight repulsion relative to 
its non-interacting counterpart. 
Solution II  represents a 
cohesive system up to the saturation point at $k_F=1.53$~fm${}^{-1}$,
where we obtain $B/A=-16.9$~MeV for its binding energy per nucleon.
Each solution leads to distinct behaviors in their 
corresponding effective masses.
While the correlated gas phase shows an increasing effective mass,
reaching $m^*\approx 3m$ near $k_\beta$,
the condensed phase starts out with near-bare-mass sp at 
$k_\alpha$, reaching a maximum of $1.8m$ at $k_F\approx 0.55$~fm${}^{-1}$,
and decreasing to about $0.8\,m$ at saturation density.
The emergence of massive sp solutions is a peculiarity
of our results, in formal resemblance to 
Heavy Fermions (electrons) observed in unconventional superconductors 
\cite{Stewart84,Gegenwart08}.
Work in progress using alternative realistic \emph{NN} interactions 
also yields coexisting solutions with large effective masses, 
confirming the robustness of these findings. 

Dinucleon structures associated 
with phases I and II for the sp fields exhibit unconventional 
spatial properties, with mean radii as large as $\sim\! 100$~fm. 
In the case of \SO pairs, their size is always
greater than the internucleon separation, with a sudden increase
in the transit from phase I to phase II.
Deuterons wavefunctions, in turn, appear very confined in phase I but 
outrange the internucleon separation in phase II.

Findings for nuclear matter in normal state were contrasted with 
those obtained from BCS pairing states, where BCS gap equations 
were solved considering sp spectra for phases I and II, independently. 
Nearly total deuteron condensate is obtained in phase I,
decreasing monotonically in phase II up to $k_F=1.8$~fm${}^{-1}$, 
where \SD pairing vanishes.
The account for superfluidity lowers, by about 6\%, 
the saturation density of nuclear matter relative to that
obtained for normal state. 
At these densities the condensation energy is much weaker than 
the sp energy, below 30\% for $k_F\gtrsim 1.2$~fm$^{-1}$,
justifying the separation between both contributions.
Such is no longer the case at low densities where condensation and
sp energies become comparable.

The disclosure of coexisting solutions in symmetric nuclear matter
at low densities constitutes a major finding in this work. 
A crucial role was played by the explicit handling of dinucleon 
bound states together with refined methods to 
obtain self-consistent solutions for the sp fields. 
We have shown that the low-density solutions for the sp potentials
$U(k)$ are not quadratic in $k$ as they exhibit a sudden change 
of slope at $k/k_F\approx 1.15$.
This feature rules out the reliability of using the effective mass
approximation for the sp fields at low densities.
In any case, the results reported here demonstrate that self-consistency
within the BHF approximation, accounting for dinucleon bound states,
is feasible without the need to assume a functional form to the 
sp fields.

We have next extended the present BHF study to include
BCS pairing to lowest order in the sp momentum distribution. 
From this extension it becomes clear that the separation 
between BHF and superfluidity energies at low densities is
not meaningfull. 
This leaves open the question as to whether the two coexisting
solutions obtained at the BHF level persist if a more complete 
treatment of superfluidity were made.
Actually, we have found that the inclusion of superfluidity
attenuates considerably differences of the binding energy of the
system in the coexisting range.
In this regard, the inclusion of hole-hole propagation needs to
be addressed to clarify whether coexisting phases correspond to
actual physical properties of nuclear matter at low densities, 
or they are a result of an oversimplified description.
The $T$-matrix approximation in self-consistent Green's
function method \cite{Bozek2002,Muther2005} offers a more
coherent framework to reassess these issues.

The study presented here, where emphasis has been placed on
pp correlations with explicit treatment of dinucleon bound states, 
is basic in the sense that multi-particle and clustering 
were not taken into account, although they are known be 
important in low-density clustering 
models \cite{Typel10,Sogo10,Horowitz06}.
Additionally, three-body forces are known to be crucial to account
for the saturation point of nuclear matter \cite{Li08}, so that their 
effect at normal as well as high densities are expected to be important. 
However, beyond such simplifying assumptions as embodied by the 
BHF framework to treat nuclear correlations, it is reasonable 
to conceive that the model retains leading-order features in 
diluted nuclear environments expected 
to linger at low temperatures and isospin asymmetry.
In this sense the findings reported in this work may provide 
additional conceptual 
tools to better understand the physics of diluted nuclear matter in the 
context of surface-sensitive nuclear reactions,
the coupling to continuum in modeling nuclear open 
quantum systems \cite{Michel10} or dense star environments.
From a more general perspective, it is also reasonable to expect 
that phase-coexistence in homogeneous nuclear matter in normal 
state breaks down as temperature and isospin asymmetry reach some 
critical values.
Identifying the conditions under which coexistence persists
constitute additional extension of this work.

\begin{acknowledgement}
H.F.A. thanks colleagues at CEA, Bruy\`eres-le-Ch\^atel, France, for their 
warm hospitality during his stay where part of this work took place.
The authors are indebted to M. Baldo and P. Schuck 
for useful discussions.
Funding from CEA/DAM/DIF is acknowledged.
This research has been funded in part by FONDECYT under grant No 1120396.
\end{acknowledgement}

\appendix
\numberwithin{equation}{section}
\section{Integral equation with multiple roots on the real axis}
\label{multiple}

A common technique to solve integral equations in momentum space
is by discretizing the momentum in the range $q:0\to\infty$, 
reducing the problem to a matrix form. Thus, the equation 
\[
t(k',k) = v(k',k) + 
\int_0^\infty \frac{F(k',q)\,dq}{\omega+i\eta-E(q)}t(q,k)\;,
\]
is reduced to the matrix equation
\begin{equation}
\label{tij}
t_{ij}=v_{ij} + \sum_{k=1}^{N} \frac{F_{ik} \,w_k}{\omega-E_k}\,t_{kj}
-i\pi\sum_{p=N+1}^{N+K} \frac{F_{ip}}{|E'(q_p)|} \,t_{pj}\;.
\end{equation}
Here we are assuming that there are $K$ solutions to  $E(q)=\omega$,
with $q$ above the minimum value allowed by Pauli blocking,
leading to the set of roots $\{ q^{(1)},\cdots,q^{(K)}\}$.
The first sum in Eq. (\ref{tij}) represents the principal-value integral,
for which we apply Gaussian-Legendre quadrature to integrate between 
consecutive roots, up to infinity.

\section{Adaptive trapezoidal quadrature}
\label{quadrature}
Consider the evaluation of the integral
\begin{equation}
\label{integral}
I\equiv\int_0^{1}f(x)dx\,,
\end{equation}
with $f(x)$ any finite function in the range $[0,1]$.
This function may also have a finite number of narrow peaks.
Let us define
\begin{equation}
\label{int0}
I_0=\textstyle{\frac12}\left [ f(0)+f(1) \right]\;.
\end{equation}
Additionally, let us consider the following sequence of sets of points 
in the interval $(0,1)$:
$s_1=\{\nicefrac{1}{2}\}$,
$s_2=\{\nicefrac{1}{4},\nicefrac{3}{4}\}$,
$s_3=\{\nicefrac{1}{8},\nicefrac{3}{8},
       \nicefrac{5}{8},\nicefrac{7}{8}\}$, etc.
Note that the union of the first $N$ sets yields a uniform distribution
of points, with no repetition of its elements. 
If the elements of the union are sequenced in ascending order, 
the separation between consecutive elements is $1/2^N$.
With this construction it can be verified that a trapezoidal
quadrature can be represented by the recurrence
\begin{equation}
\label{recurrencia}
I_n=\textstyle{\frac12} I_{n-1}+\textstyle{\frac{1}{2^n}}\sum_{x_j\in s_n}f(x_j)\;,
\end{equation}
where $I_0$ is given by Eq. (\ref{int0}). 
In actual evaluations the iteration is interrupted at some $n=M$ when a 
convergence criteria for $|I_M-I_{M-1}|$ is met. 
If narrow peaks occur, the minimum $n$ should be such that
$1/2^n$ becomes smaller than the smallest width.
The advantage of this method is that the historical reckoning of $f(x)$
is not discarded if a narrower mesh is needed.
This is an important consideration whenever a single evaluation of $f$ 
is time-consuming.

\section{Evaluation of $\Psi_0$ for S and D waves}
\label{exactfourier}
Let us define 
\begin{equation}
\label{phi0}
\psi_0^{(L)}(q) = A q^L\exp[-R(q-\bar q)]\Theta(q-\bar q)\;.
\end{equation}
Denoting $z=\bar qR$, and $x=\bar q r$, 
the Fourier transform for $\Psi_0^{(L)}$ for S waves becomes 
\begin{align}
\label{fourierS}
\Psi_0^{(0)}(r) = &
\sqrt{\frac{2}{\pi}}\,
\frac{A\bar q^{3}}{x(z^2 + x^2)^2} 
\left [
  (2z + z^2 + x^2)x \cos x  + \right . \nonumber \\
&
\left. (z^2 + z^3 + zx^2 - x^2) \sin x 
\right ]\;.
\end{align}
The asymptotic behavior $r\to\infty$ is identified for $x\gg 1$,
case in which we get
\begin{equation}
\label{fourierSlarge}
\Psi_0^{(0)}(r) \approx
A\bar q\;
\sqrt{\frac{2}{\pi}} \;
\frac{\cos(\bar qr)}{r^2}\;.
\end{equation}

In the case of D-waves we obtain
\begin{align}
     \Psi_0^{(2)}(r)=&
\sqrt{\frac{2}{\pi}}\,
\frac{-A\bar q^5}{x^3(x^2+z^2)^4}  
 \left \{
 \left [ x^8+3 x^6 \left(z^2+3 z-5\right) +
 \right .
 \right . 
 \nonumber \\
  &  3 x^4 z \left(z^3+7 z^2-z-16\right) +
       x^2 z^4 \left(z^2+15 z+15\right) + 
\nonumber \\
  &3 z^6 (z+1) ] x \cos(x) + \left [ x^8 (z - 6) - 3 z^6 (z + 1) +\right.
\nonumber \\
  &  x^2 z^4 (z^3 - 15 z - 15) + 3 x^6 (z^3 - 4 z^2 - 11 z + 5) + 
\nonumber \\
  & \left .  
      3 x^4 z^2 (z^3 - 2 z^2 - 15 z - 15)
      \right ] \sin(x)   \} \;.
\end{align}
This result yields for large distances ($x\gg 1$)
\[
\Psi_0^{(2)}(r) \approx
-A\bar q^3\; \sqrt{\frac{2}{\pi}} \;
\frac{\cos(\bar qr)}{r^2}\;.
\]
%
%
%
%

\end{document}